\documentclass[11pt]{article}
\usepackage[utf8]{inputenc}

\usepackage{stmaryrd}
\usepackage{varioref}
\usepackage{amsmath,amsfonts,amsthm,mathrsfs}
\usepackage{mathabx}
\usepackage[normalem]{ulem}
\usepackage{multirow,color,graphics}
\usepackage{amsmath}
\usepackage{amsfonts}
\usepackage{amssymb}
\usepackage{jheppub}
\usepackage{enumerate}
\usepackage{fancyvrb}
\usepackage{verbatim}
\usepackage{wrapfig}
\usepackage{appendix}
\usepackage{amstext}
\usepackage{amssymb}
\usepackage[bbgreekl]{mathbbol}
\usepackage{bbm}
\usepackage{graphicx}
\usepackage{color}
\usepackage{xcolor}
\usepackage{varioref}
\usepackage{multirow,graphics}
\usepackage{epstopdf}

\usepackage{mathbbol}

\DeclareSymbolFontAlphabet{\mathbbm}{bbold}
\DeclareSymbolFontAlphabet{\mathbb}{AMSb}

\renewcommand{\comment}[1]{}

\numberwithin{equation}{section}

\def\[{\left[}
\def\]{\right]}
\def\({\left(}
\def\){\right)}
\def\<{\left<}
\def\>{\right>}

    \newcommand{\beq}{\begin{equation}}
    \newcommand{\eeq}{\end{equation}}
    \newcommand\beqa{\begin{eqnarray}}
    \newcommand\eeqa{\end{eqnarray}}
\newcommand\bea{\begin{array}}
\newcommand\eea{\end{array}}

\newcommand{\bQ}{{\bf Q}}

\newcommand{\bP}{{\bf P}}

\newcommand{\la}[1]{\label{#1}}
\newcommand{\eq}[1]{(\ref{#1})}

    \def\bQ{{\bf Q}}
    \def\bP{{\bf P}}

\definecolor{cadmiumgreen}{rgb}{0.0, 0.42, 0.24}

\makeatletter
     \@ifundefined{usebibtex}{} {}
\makeatother

    \def\bQ{{\bf Q}}

        \def\bP{{\bf P}}

     \def\Tr{\text{Tr}}

\newcommand{\Ptt}{\mathbb{P}} 

\renewcommand{\<}{\langle} 
\renewcommand{\>}{\rangle} 
\renewcommand{\sl}{\mathfrak{sl}}

\newcommand{\lO}{\mathcal{O}}
\newcommand{\lN}{\mathcal{N}}

\newcommand{\ii}{i}

\usepackage{dsfont}

\newcommand{\AdSST}{AdS$_3 \times $ S$^3 \times $ T$^4$ }
\newcommand{\parityg}{\mathbf{g}}

\newcommand{\aparam}{d}

\title{
Exploring the Quantum Spectral Curve for AdS$_3$/CFT$_2$
}

\emailAdd{andrea.cavaglia$\bullet$unito.it} 
\emailAdd{simon.ekhammar$\bullet$physics.uu.se}
\emailAdd{nikgromov$\bullet$gmail.com}
\emailAdd{paul.1.ryan$\bullet$kcl.ac.uk} 

\author[a]{Andrea Cavagli\`a}

\author[b]{Simon Ekhammar}

\author[c,d]{Nikolay Gromov}

\author[c]{Paul Ryan}

\affiliation[a]{
Department of Physics, University of  Turin, Via P. Giuria 1, 10125, Turin, Italy
}

\affiliation[b]{
Department of Physics and Astronomy,
Uppsala University, Box 516, SE-751 20 Uppsala, Sweden
}

\affiliation[c]{
Mathematics Department, King's College London,
The Strand, London WC2R 2LS, UK
}

\affiliation[d]{St.Petersburg INP, Gatchina, 188 300, St.Petersburg,
Russia}

\abstract{
Despite the rich and fruitful history of the integrability approach to string theory on the \AdSST background, it has not been possible to extract many concrete predictions from integrability, except in a strict asymptotic regime of large quantum numbers, due to the severity of wrapping effects. 
The situation changed radically with two independent and identical proposals for the Quantum Spectral Curve (QSC) for this system in a background of pure Ramond-Ramond flux. This formulation is expected to capture all wrapping effects exactly and describe the full planar spectrum. 
Massless modes conjecturally manifest themselves in a new property of this QSC: the non-quadratic nature of the branch-cut singularities of the QSC Q-functions. This feature implies new  technical challenges in solving the QSC equations as compared  to the well-studied case of ${\cal N}=4$ SYM. In this paper we resolve these difficulties and obtain the first ever predictions for generic unprotected string excitations. We explain how to extract a systematic expansion around the analogue of the weak 't Hooft coupling limit in $\mathcal{N}$=4 SYM and also obtain high-precision numerical results. This concrete data and others obtainable from the QSC could help to identify the so-far mysterious dual CFT. 
}

\begin{document}
\maketitle

\section{Introduction}
AdS$_3$/CFT$_2$ dualities realise the AdS/CFT correspondence~\cite{Maldacena:1997re} in a seemingly simplified setting, due to the lower degrees of freedom on the gravity side  and the special features of 2D CFTs. Thus, they offer ideal prototype examples, where the mechanisms of holography should be simpler to understand. The price to pay is that concrete AdS$_3$/CFT$_2$ dualities are in general  hard to define precisely  and to study quantitatively. In particular, in all known examples the dual CFTs lack a Lagrangian description when generic  interactions are switched on. This feature, together with their rich moduli spaces with many more parameters than in higher dimensional analogues, makes these dualities fascinating and at the same time hard to  investigate, with the exception of special limits where simplifications occur.\footnote{
A notable case is the tensionless limit of string theory on ${\rm AdS}_3\times {\rm S}^3\times {\rm T}^4$ with one unit of NS-NS flux, which was argued to be exactly dual to the free symmetric product orbifold CFT~\cite{Eberhardt:2018ouy,Eberhardt:2019ywk} (for a concrete proposal with more general NS-NS flux see \cite{Eberhardt:2019qcl}), and is essentially solvable even outside the planar limit. This theory may be seen as an analogue of free $\mathcal{N}=4$ SYM in the case of AdS$_5$/CFT$_4$. Moving beyond this point is where we believe integrability and the techniques of this paper play an important role.
}

\paragraph{${\rm AdS}_3\times {\rm S}^3 \times {\rm T}^4$ with pure RR flux.}

  In this paper we study an important example of AdS$_3$/CFT$_2$~\cite{Maldacena:1997re}, which until now was almost impenetrable to computations of non-protected quantities. 
  It is defined, on the gravity side, by string theory on an \AdSST background supported by pure Ramond-Ramond (R-R) flux. The dual CFT arises as an IR limit of the worldvolume gauge theory of the D1-D5 brane system. 
  Not much is known about this CFT, except that it is expected to live on the same conformal manifold as the symmetric product orbifold sigma model (arguments for this, and further speculations were made e.g. in \cite{Maldacena:1997re,Seiberg:1999xz,Larsen:1999uk,Pakman:2009mi,OhlssonSax:2014jtq}). At the same time, worldsheet CFT methods are prohibitively hard to use with R-R fluxes.\footnote{This is opposed to the case of NS-NS flux, where there is an explicit WZW description of the worldsheet CFT~\cite{Maldacena:2000hw}, making it far more tractable with traditional methods than the R-R case.}  
  There is however compelling evidence that string theory on this background  exhibits quantum integrability on the worldsheet~\cite{Babichenko:2009dk} (for a review see e.g. \cite{Sfondrini:2014via}, and \cite{OhlssonSax:2011ms,Borsato:2014exa,Borsato:2014hja,Borsato:2016kbm,Borsato:2016xns} for developments related to the model considered here). 
  Thus, integrability gives the best opportunity to solve the theory. 
  In this paper we show a new integrability-based tool in action, the Quantum Spectral Curve (QSC) proposed simultaneously in ~\cite{Ekhammar:2021pys} and \cite{Cavaglia:2021eqr}, which was conjectured in these works to encode the full planar spectrum in the sector with zero winding and momentum on ${\rm T^4}$. Solving these equations, we obtain the first predictions for non-protected scaling dimensions of the dual CFT at finite values of the string tension (which plays the role of a non-perturbative coupling constant).
  
\paragraph{Integrability in AdS/CFT.}
  Historically, integrability was first discovered for string theory on AdS$_5\times $S$^5$, dual to $\mathcal{N}$=4 SYM theory, and shortly after for AdS$_4\times  $CP$^3$, dual to ABJM theory (for reviews, see \cite{Beisert:2010jr,Gromov:2017blm}). 
  It is also expected that integrability persists for several AdS$_3$ backgrounds, which includes AdS$_3\times $S$^3\times $T$^4$ and AdS$_3\times $S$^3\times $S$^3\times $S$^1$ with several combinations of fluxes. It is useful to recall how the spectral problem was solved in the AdS$_5$ and AdS$_4$ cases, to highlight some important differences pertaining to AdS$_3$. 
The route to solving the spectrum starts from the understanding of the dynamics of excitations on a very large string worldsheet of size $L$ in lightcone gauge. In this gauge, worldsheet particles are  massive in the AdS$_5$ and AdS$_4$ cases, while for the AdS$_3$ case they also include massless modes. These massless modes increase the sheer complexity of the system, but also lead to subtle effects. 
  In the large-$L$ regime, the spectrum of these integrable systems is described by a set of Asymptotic Bethe Ansatz (ABA) equations. These equations are valid up to corrections known as \emph{wrapping effects}. While in a massive integrable system  these corrections are exponentially suppressed with $L$, in the presence of massless modes as in the AdS$_3$ case they already kick in at $\mathcal{O}(1/L)$ order~\cite{Abbott:2015pps,Abbott:2020jaa}.
  
  The previously mentioned Quantum Spectral Curve formalism takes wrapping effects completely into account. This was first developed for AdS$_5$~\cite{Gromov:2013pga,Gromov:2014caa} and then AdS$_4$~\cite{Cavaglia:2014exa,Bombardelli:2017vhk},
   and recently conjectured for the AdS$_3$ duality we are considering here~\cite{Cavaglia:2021eqr,Ekhammar:2021pys}. 
   In the AdS$_5$  and AdS$_4$ cases, the QSC has revolutionised the computational possibilities, allowing e.g. for numerical studies of the spectrum with essentially arbitrary precision~\cite{Gromov:2015wca}, as well as exact analytic results in the gauge theory weak coupling expansion and other regimes~\cite{Marboe:2014gma,Gromov:2015vua}. 
   
   Moreover, the QSC is at the core of the integrable structure of the theory, and various approaches connect it to the computation of correlation functions. This is suggested by the fact that the solutions of the QSC should provide building blocks for the Separation of Variables, see e.g.~\cite{Cavaglia:2018lxi,Giombi:2018hsx,Cavaglia:2021mft,Bercini:2022jxo} for some examples. Furthermore, a recent proposal for the exact form of some OPE coefficients  in the full $\mathcal{N}$=4 SYM theory  links them to the solutions of the QSC~\cite{Basso:2022nny}. 
  
  On top of this, one can use a hybrid approach known as Bootstrability by feeding high-precision spectral data from the QSC into the conformal bootstrap, which has been shown to very precisely constrain some OPE coefficients~\cite{Cavaglia:2021bnz,Cavaglia:2022qpg,Caron-Huot:2022sdy}. 
  All these ongoing developments provide further strong motivation to study the QSC for this AdS$_3$/CFT$_2$ duality, where we  expect it will lead to applications beyond the spectrum (see also \cite{Eden:2021xhe,Fabri:2022aup} for progress from the complementary hexagon approach of \cite{Basso:2015zoa}).

\paragraph{Quantum Spectral Curve.}
  Previous examples of QSCs for AdS$_5$ and AdS$_4$  were deduced  starting from the ABA, going through the thermodynamic Bethe ansatz (TBA) -- a procedure to resum wrapping effects -- and then going through a long and complex chain of simplifications. TBA equations for AdS$_3$ have recently been proposed~\cite{Frolov:2021bwp}, after amending a proposal for the form of the dressing phases of the worldsheet S-matrix~\cite{Frolov:2021fmj}.
  Even though the TBA equations use a number of additional assumptions,  this process could in principle now be repeated to provide an independent test for the QSC proposal of \cite{Ekhammar:2021pys,Cavaglia:2021eqr}.   
  
  The QSC papers \cite{Ekhammar:2021pys,Cavaglia:2021eqr} instead used a new and less orthodox approach, by considering a classification of the possible QSC  mathematical structures. This method used only the global symmetry of the system as the input, as well as  minimal analyticity assumptions coming from the form of the ABA equations. This was found to lead to a natural proposal for the QSC. Given this non-standard approach, it is clearly very important to test the  conjecture.\footnote{We emphasise however that even the `canonical' argument to deduce the QSC from TBA is based on (at least) two different kinds of conjectures. First of all, the TBA is based on the assumption of quantum  integrability, of which there is no rigorous proof for the AdS/CFT models. At a more technical level,  the TBA uses  the detailed form of the worldsheet S-matrix, which in principle suffers from ambiguity of the CDD factors in the dressing phases. With respect to this second point, the argument of \cite{Ekhammar:2021pys,Cavaglia:2021eqr} may be more robust, since it does not use the details of the S-matrix.
  } A consistency check performed in \cite{Ekhammar:2021pys,Cavaglia:2021eqr} showed that, in the limit of large volume, the QSC equations reproduce the result of the ABA, at least for the infinite ``massive'' subsector.\footnote{The states in this sector are described at large volume by a gas of massive worldsheet particles. At finite volume, however, they receive  contributions from both massive and massless \emph{virtual} particles.} While the asymptotic analysis is trickier, it  was conjectured in the same works that the QSC describes the full spectrum including all asymptotically massless modes. 
  
  In this paper, we show for the first time that the new QSC equations have nontrivial isolated solutions, which is by no means guaranteed by the initial abstract construction based on  symmetries. We view this as an additional strong argument towards the validity  of the original QSC conjectures.
  To do this, we have to deal with a surprising new feature of the QSC, which makes the solution of the equations more challenging. Namely, while in previous cases the analytic properties of the QSC were described in terms of quadratic branch cuts in the spectral parameter domain, in the AdS$_3$ case the cuts have in general infinite order. This new property emerged out of the self-consistency of the formalism, and it was proposed in \cite{Ekhammar:2021pys,Cavaglia:2021eqr} that it is in fact a signature of the presence of massless modes. 
  This new feature forces us to drastically modify the previous methods to solve the QSC equations. 
  
  In  this paper we resolve these difficulties and  develop new general methods, which open the way to an in-depth study of the spectrum. We focus  on the regime of finite string tension, and the analytic expansion at weak ``coupling" constant $g \sim 0$ (the analog of $g=\frac{\sqrt{\lambda}}{4\pi}$ in $\mathcal{N}=4$ SYM and $h(\lambda)$ in ABJM theory), where we compute up to $g^8$ order (finding the first $7$ non-trivial coefficients analytically). The coupling $g$ (called ``$h$'' in \cite{Ekhammar:2021pys,Cavaglia:2021eqr}) enters into the integrability formulation via the position of the branch cuts and is related to the string tension $T = \frac{1}{2 \pi \alpha'}$ as $g \sim R_{AdS}^2/\alpha'$ in the regime $g\to\infty$ .\footnote{In general, it is expected that $g$ will be a function of additional moduli of the theory on top of $\alpha'$, see \cite{OhlssonSax:2018hgc}.} 
  From now on, we simply refer to $g$ as ``the coupling constant''.  
  In our case, we do not have a Lagrangian, and the $g\sim 0$ region we study is not yet accessible with any other known method. 
  
  \paragraph{Domain of validity and assumptions.} 
  The QSC equations we use in this paper were conjectured in \cite{Ekhammar:2021pys,Cavaglia:2021eqr} based on symmetries of the ${\rm AdS}_3 \times {\rm S}^3 \times {\rm T}^4$ model in the sector with zero winding and momentum on the ${\rm T}^4$, which we restrict to. In those works, the QSC equations are shown to be consistent with the known ABA equations.

  In this paper we also discuss another important ingredient of the QSC construction -- the constant gluing matrix $G$. We will see that the analytic properties of the Q-functions depend significantly on the form of $G$, and we identify the most natural choice which also leads to maximal analyticity of the Q-function. Even though we give strong evidence towards the validity of this point, this still remains an additional assumption in our construction.

  We emphasise that the results of this paper are the first ever predictions for generic unprotected string excitations. Assuming the correctness of the QSC, such precise data should help to test conjectures on the form of the dual CFT$_2$~\cite{Maldacena:1997re,Seiberg:1999xz,Larsen:1999uk,Pakman:2009mi,OhlssonSax:2014jtq}.

\paragraph{Structure of the paper.}  
 The rest of the paper is organised as follows.  In section \ref{sec:QSCreview}, 
 we review the proposal for the QSC, while in section \ref{sec:Ptt}
 we introduce new tools needed to deal with the new non-quadratic branch points. In  section \ref{sec:numerics}, we discuss a method to solve the equations numerically at finite coupling, contrasting it with the previous algorithm in AdS$_5$. In section \ref{sec:perturbative}, we discuss the perturbative method at weak coupling and list our results in this regime. Finally, section \ref{sec:discussion} contains a discussion of the results and possible future directions. The paper is closed by two appendices which present additional  aspects of the QSC, and unpack some technical details of the weak coupling expansion.
 
\section{Review of the QSC for ${\rm AdS}_3\times {\rm S}^3\times {\rm T}^4$}\label{sec:QSCreview}
In this section we review the original QSC construction for AdS$_3$ and introduce notations needed in the rest of the paper, where we will introduce additional tools and explain how to solve the equations numerically and analytically.

The symmetry algebra of the ${\rm AdS}_3 \times {\rm S}^3 \times {\rm T}^4$ system in the sector with zero winding and momentum on the torus is $\mathfrak{psu}(1,1|2)^{\oplus 2}$, with the bosonic subalgebra $ \left(\mathfrak{su}(1,1)\times \mathfrak{su}(2) \right)^{\oplus 2}$. 
Correspondingly, we introduce $4$ sets of  indices running from 1 to 2 to label vectors covariant with respect to these subalgebras. 
The Roman letters from the start of the alphabet $a,b,\dots$ take values $1$ or $2$, and similarly their dotted versions $\dot{a},\dot{b},\dots$ take values $\dot{1},\dot{2}$. These indices represent the two $\mathfrak{su}(2)$ subalgebras. The other two sets of indices are the Roman letters from the middle of the alphabet $k,l,\dots$
and $\dot{k},\dot{l},\dots$, which are associated to the two $\mathfrak{su}(1,1)$ subalgebras. 
Summation over repeated indices is always assumed, unless explicitly stated otherwise. 

\subsection{$\bP\mu$-system}

The quantum spectral curve of the ${\rm AdS}_3\times {\rm S}^3\times {\rm T}^4$ integrable system is most conveniently described in terms of the $\bP\mu$ system. This is a set of functional relations between  $\bP$ functions: $\bP_a(u)$, $\bP^a(u)$ and $\bP_{\dot{a}}(u)$, $\bP^{\dot{a}}(u)$ and $\mu$ functions: $\mu_a^{\ \,\dot{b}}(u)$ and $\mu_{\dot{a}}^{\ \, b}(u)$. To lighten the notation we will sometimes write $\mu$ and $\dot \mu$ for these two matrices. Moreover, we have $\text{det}(\mu) = \text{det}(\dot \mu)=1$~\cite{Ekhammar:2021pys,Cavaglia:2021eqr}. 

To simplify exposition we will often make statements for functions with undotted indices, with the understanding that the same relation holds after swapping all undotted indices with their dotted versions and vice-versa.  From  $\bP_{a}\bP^{a}=0$ it follows that:\footnote{The Levi-Civita symbol is defined by $\epsilon_{12}=-\epsilon_{21} = 1$, $\epsilon_{11}=\epsilon_{22} = 0$, and $\epsilon_{ab} \epsilon^{bc} = -\delta_a^c$. }
\beq\la{ul}
\bP_a= r \epsilon_{ab} \bP^b ,
\eeq
 where $r(u)$ is a simple rational function of the Zhukovsky variable $x(u)$  defined by\footnote{The Zhukovsky variable has sheets on its Riemann surface where it has long or short cuts. We choose the sheet with a short cut.}
\begin{equation}\la{eqZ}
    x(u) = \frac{u}{2g}+\sqrt{\frac{u}{2g}+1}\sqrt{\frac{u}{2g}-1}\;.
\end{equation}
Later on we will specialise to a specific subsector of states, which we call the $\sl(2)$-sector, where we simply have $r(u) = \dot r(u) = 1$. 
 
The functions $\bP_a$ carry $a$-type of indices and can be associated with S$^3$ degrees of freedom. They have branch points at $u=\pm 2g$, which are connected by a short cut, and have no other singularities. We refer to this sheet of the Riemann surface of $\bP_a$ as the \textit{defining sheet}. On the other hand, $\mu_a^{\ \,\dot{b}}$ have an infinite ladder of short cuts connecting branch points located at $\pm 2g+ n\,i$, $n\in \mathbb{Z}$.

\begin{figure}[t]
    \centering
    \includegraphics[scale=0.2]{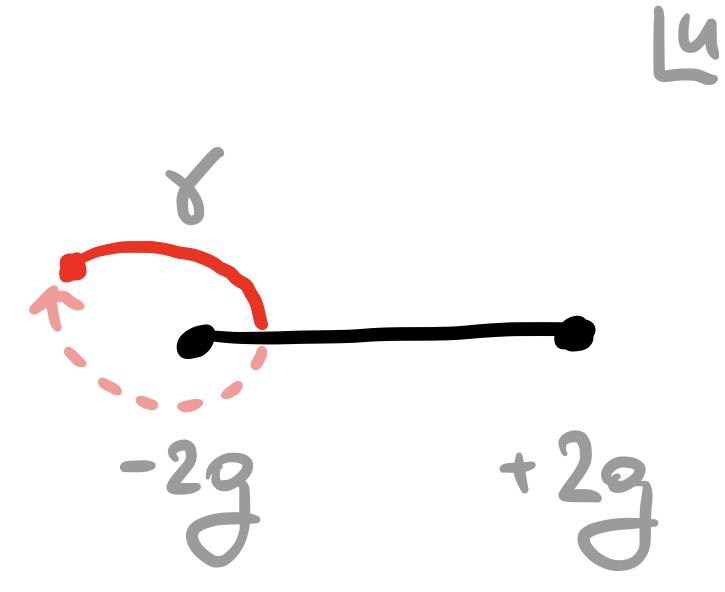}\;\;\;\;\;\;\;\;\;\;\;\;\;\;\;
        \includegraphics[scale=0.2]{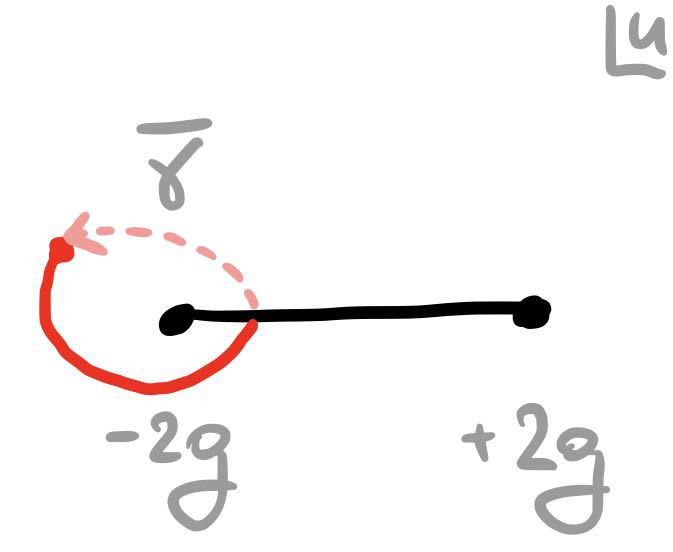}
    \caption{Two non-identical analytic continuations around the branch point at $u=-2g$ we denote by $\gamma$ (clock-wise) and $\bar\gamma$ (anti-clock-wise).}
    \label{fig:gamma}
\end{figure}
\paragraph{Analytic continuation.}
In contrast to the well-studied ${\rm AdS}_5$ and ${\rm AdS}_4$ integrable models the branch points entering the $\bP\mu$ system of the ${\rm AdS}_3$ model are generically not quadratic. This means that the result of analytically continuing these functions around the branch points in the clockwise 
or anti-clockwise direction will yield different results. For a function $f$ with a short cut $[-2g,2g]$ we use $f^\gamma$ ($f^{\bar{\gamma}})$ to denote the analytic continuation of $f$ around the branch point $-2g$ in the clockwise (counterclockwise) direction. We summarise this prescription for analytic continuation in Figure \ref{fig:gamma}.

The $\bP\mu$-system equations can now be stated as 
\begin{equation}\la{Pgamma}
    (\bP_a)^{\bar{\gamma}}=\bP_{\dot{b}}\,\mu^{\dot{b}}_{\;\; a}\;\;,\;\;(\mu_a^{\;\;\dot b})^\gamma-\mu_a^{\;\;\dot b}= \bP_a (\bP^{\dot b})^{\bar{\gamma}} - (\bP_a )^{\gamma} \bP^{\dot b} ,
\end{equation}
where $\mu^a_{\;\;\dot b}$ denotes the inverse transpose of $\mu$: $\mu^a_{\;\;\dot b} = -\epsilon^{ab} \mu_b^{\;\;\dot a} \epsilon_{\dot a \dot b}$ and $\mu_a^{\;\;\dot b}\mu^c_{\;\;\dot b}=\delta_{a}^{\;\;c}$. These equations should be supplemented by large $u$ asymptotics, described in detail below in subsection 
\ref{sec:sl2sec}, and also by a crucial analytic property for the $\mu$ functions: 
they are ``mirror-periodic''\footnote{We use standard notation for shifts of the 
argument of a function $f(u)$: $f^\pm :=f(u\pm\frac{i}{2})$, $f^{\pm\pm} :=f(u\pm i)$, $f^{[n]}:=f(u+i\frac{n}{2})$, $n\in\mathbb{Z}$. This shifts are implied to be on the Riemann section with short cuts.}
\beq
\mu^{\gamma} = \mu^{++},
\eeq
which means that they are $i$-periodic when shifted on a special ``mirror'' section of the Riemann surface defined with long cuts~\cite{Ekhammar:2021pys,Cavaglia:2021eqr}.

As shown in \eqref{Pgamma}, the analytic continuation of the $\bP$ functions is governed by the matrix $\mu$. As a result of the fact that in AdS$_3$ the cuts are no longer quadratic as in previous cases, it is  convenient to introduce a ``regularized" version of $\mu$-functions $\mu^R$, which has the cut on the real axis ``removed"
\begin{equation}\label{muRdef}
    \mu_a^{R\; \dot{b}} = (\delta_{a}^{\;\;b} + \bP_a\bP^b)\mu_b^{\;\;\dot{b}}\,.
\end{equation}
This is helpful because using $\bP_a\bP^a=0$ we can replace $\mu$ with $\mu^R$ in \eq{Pgamma}. Since $\mu^{R}$ does not have a cut on the real axis, it follows that one can apply analytic continuation of $\bP$ along $\gamma$ multiple times, each time simply multiplying by the corresponding $\mu^R$ or $\dot{\mu}^R$.
In particular the double $\gamma$ continuation is given in terms of 
\beq\la{Wdef}
W=\mu^R\dot{\mu}^R\;,
\eeq
such that
\begin{equation}\label{Pdoublegamma}
    (\bP_a)^{2\gamma} = W_a^{\;\; b}\bP_b\,.
\end{equation}
In the case of AdS${}_5$, the analog of the matrix $W$ satisfies $W=1$ and all the cuts are quadratic as a consequence. In the current case, a priori there is no reason to assume $W=1$ as discussed in detail  in~\cite{Ekhammar:2021pys,Cavaglia:2021eqr}. 

In the next section we will explain how to compute $\mu^R$ from a given set of $\bP$ functions via the QSC QQ-system, which will allow us to close the system of equations.

\subsection{$\bQ$-functions and gluing conditions}
The $\bP\mu$-system described in the previous section is a consequence of the QQ-relations (which follow from the symmetry of the problem) and the gluing conditions imposed on a particular subset of Q-functions: namely, those Q-functions with indices representing the $SU(1,1)^2$ symmetry subgroup associated with ${\rm AdS}_3$. These functions are called ${\bf Q}_k$ and ${\bf Q}_{\dot k}$, and we describe the gluing conditions they satisfy in this section.

The functions $\bQ_k$ are not independent from the $\bP$-functions, and can in fact be restored from known $\bP$'s as solutions of the following second-order Baxter equation 
\begin{equation}\la{bax}
    \bQ_k^{++}D_1^- - \bQ_k D_2 + \bQ_k^{--}D_1^+=0\,,
\end{equation}
where the coefficients are given in terms of $\bP$'s as follows 
\begin{equation}
    \begin{split}
        & D_1 = \epsilon_{ab}(\bP^{a})^-(\bP^{b})^+ ,\\
        & D_2 = \epsilon_{ab}(\bP^{a})^{--}(\bP^{b})^{++}-\bP_{c}(\bP^{c})^{--}\, \epsilon_{ab}\bP^{a}(\bP^{b})^{++} \,.
    \end{split}
\end{equation}
On the defining sheet of the $\bP$-functions, $\bQ_k$ have an infinite ladder of short cuts as can be seen from the Baxter equation. Being a finite-difference equation of order two, the latter has two linearly independent solutions $\bQ_k$, $k=1,2$. However, there is a freedom in how to choose the two linearly independent $\bQ_k$. One possible choice is to take solutions to be upper half plane analytic (UHPA), with an infinite ladder of cuts in the lower-half plane. We denote this choice by $\bQ^\downarrow_k$. In combination with the large $u$ asymptotics, this fixes $\bQ^\downarrow_k$'s uniquely. Another option is to take the $\bQ_k$ to be lower-half plane analytic (LHPA) with an infinite ladder of cuts in the upper-half plane, which we denote $\bQ^\uparrow_k$. 

As explained in~\cite{Ekhammar:2021pys,Cavaglia:2021eqr}, the additional gluing condition which one has to impose, in analogy with the AdS$_{5}$ and AdS$_{4}$ cases, is that the analytic continuation of these two sets of $\bQ$'s are related by a constant ``gluing" matrix $G$
\begin{equation}
    (\bQ_k^\downarrow)^\gamma = G_{k}^{\;\; \dot{m}}\bQ_{\dot{m}}^\uparrow
\end{equation}
where we have $\det\;G=1$. 
We will additionally assume that the diagonal elements of the gluing matrix are zero ~\cite{Cavaglia:2021eqr}\footnote{This statement is only meaningful once the basis of $\bQ$s is fixed by their asymptotics like in~\cite{Cavaglia:2021eqr}.}. We will motivate this assumption at the end of this section.

In the next section we specialise to a particular sub-sector of operators and give more details on the properties of the gluing matrix in this case.

\subsection{$\sl(2)$-sector}\label{sec:sl2sec}
In this paper we specialise to a particular subset of states, which is reminiscent of the $\sl(2)$-sector of $\lN=4$ SYM. In the latter theory, the $\sl(2)$-sector corresponds to single-trace local operators $\lO$ of the form 
\begin{equation}
    \lO ={\rm Tr}\left(D^S Z^L \right)+\dots
\end{equation}
where $Z$ represents a scalar field, $D$ a lightcone derivative and $\dots$ refers to permutations. Restricting to such states produces a large simplification in the properties of the $\bP$ functions entering the QSC equations. As the dual CFT of the ${\rm AdS}_3$ model under study is not explicitly known, the $\sl(2)$ sector we consider here does not have a known realisation in terms of single-trace local operators. On the other hand, the properties of the $\bP$ functions we consider here are similar to those of the $\sl(2)$ sector of $\lN=4$ SYM and hence we will refer to the subsector of states we consider as the $\sl(2)$ sector.

A key feature of the QSC formulation is that the quantum numbers describing a given state, in particular the conformal dimension $\Delta$, are encoded in the asymptotics of the $\bP$ and $\mu$ functions at large real values of the spectral parameter $u$. In the $\sl(2)$ sector we focus on in this work the asymptotics are given by 
\begin{equation}\la{ppow}
    \bP_a \simeq A_a u^{M_a},\quad \bP_{\dot a} \simeq A_{\dot a} u^{M_{\dot a}}
\end{equation}
with 
\begin{equation}\label{Mnumbers}
    M_a = \left(-\frac{L}{2}-1,\frac{L}{2} \right),\quad M_{\dot{a}} = \left(-\frac{L}{2},\frac{L}{2}-1 \right)\;
\end{equation}
and the prefactors $A$ satisfy the relations\footnote{Note that the parameter $S$ here is related to the one in \cite{Cavaglia:2021eqr} by a change of overall sign.}
\begin{equation}\label{eq:AA}
    A_1 A^1 = -A_2 A^2 = \frac{i}{4}\frac{(\Delta-L-S)(\Delta+L-S+2)}{L+1},
\end{equation}
\begin{equation}\label{eq:AAd}
    A_{\dot{1}} A^{\dot{1}} = -A_{\dot{2}} A^{\dot{2}} = \frac{i}{4}\frac{(\Delta-L+S)(\Delta+L+S-2)}{L-1}\,.
\end{equation}
As a consequence the asymptotics of $\bQ_i$ are also fixed to
\beqa\la{qpow}
\bQ_k^\downarrow\simeq B_k u^{\hat M_k}\;\;,\;\;
\bQ_{\dot k}^\downarrow\simeq B_{\dot k} u^{\hat M_{\dot k}}
\eeqa
with
\begin{equation}\la{asm}
    \hat M_k = \left(\frac{\gamma}{2}+\frac{L}{2},-\frac{\gamma}{2}-\frac{L}{2}-1 \right),\quad 
    \hat M_{\dot{k}} = \left(\frac{\gamma}{2}+\frac{L}{2}-1+S,-\frac{\gamma}{2}-\frac{L}{2}-S \right)\;.
\end{equation}
Above, the only non-integer quantum number is $\gamma$ -- the anomalous dimension, related to the full conformal dimension $\Delta$ as $\Delta = L + S + \gamma$. As well as this, in this subsector the function $r$ of \eqref{ul} relating $\bP$ functions with upper and lower indices is simply given by $r=1$\footnote{This follows from the explicit expression of $r$ in terms of the quantum numbers of a state, see \cite{Cavaglia:2021eqr}.} and thus $\bP^a$ and $\bP_a$ are related by
\begin{equation}
    \bP^a = -\varepsilon^{ab}\bP_b ,
\end{equation}
and similarly for the dotted $\bP$ functions. 
\paragraph{Parity symmetry.}
In addition to the $\sl(2)$ sector we consider parity invariant states, symmetric under $u\to-u$. This is for example the case for the Konishi operator of $\lN=4$ SYM. In practice this implies that the $\bP$-functions are either even or odd, depending on their asymptotics. As such we have
\begin{equation}\label{eq:ParityP}
    \bP_a(-u) = \parityg_a^{\;\; b}\bP_b(u),
\end{equation}
where $\parityg_{a}{}^{b}$ is a diagonal $2\times 2$ matrix with entries $\pm 1$. The parity structure also simplifies the gluing procedure relating the two sets of $\bQ_i$ functions $\bQ_i^\downarrow$ and $\bQ_i^\uparrow$, as these are now related by flipping the sign of $u$:
\beq
\bQ_k^\uparrow(u)=e^{-i \pi \hat M_k}\bQ_k^{\downarrow}(-u),
\eeq
where as usual the same relation remains true after interchanging dotted and undotted indices. Note that there is no summation over repeated indices on the r.h.s. The phase factor can be deduced by comparing the large $u$ asymptotics (which is valid away from the ladders of cuts) of the l.h.s. and the r.h.s.
As a consequence, the gluing conditions $\bQ_k^\downarrow(u+i0)=G_k^{\;\;\dot k}\bQ_{\dot k}^\uparrow(u-i0)$, and their dotted version, become
\beq
\bQ_k^{\downarrow}(u+i0)=G_k^{\;\;\dot k}e^{-i\pi\hat M_{\dot k}}\bQ_{\dot k}^\downarrow(-u+i0)\;\;,\;\;
\bQ_{\dot k}^{\downarrow}(u+i0)=G_{\dot k}^{\;\; k}e^{-i\pi\hat M_{ k}}\bQ_{k}^\downarrow(-u+i0)\;.
\eeq
Changing $u$ to $-u$ in the second equation and combining it with the first we get
\beq
\bQ_k^{\downarrow}(u+i0)=G_k^{\;\;\dot k}
e^{-i \pi \hat M_{\dot k}}
G_{\dot k}^{\;\; l}e^{-i\pi \hat M_l}\bQ_{l}^\downarrow(u+i0)\;,
\eeq
As this holds for any $u$ we conclude that
\beq\label{eqn:GGdotreln}
G_k^{\;\;\dot k}
e^{-i \pi \hat M_{\dot k}}
G_{\dot k}^{\;\; l}e^{-i\pi \hat M_l}=\delta_{k}^l\;.
\eeq
Assuming that the diagonal elements are zero, and recalling that $\det G=1$, we arrive at
\beq\la{Galpha}
G_{k}^{\;\;\dot k}=\(
\bea{cc}
0& i\alpha\\
i/\alpha& 0
\eea
\),
\eeq
and $G_{\dot{k}}^{\;\; k}$ can be deduced from \eqref{eqn:GGdotreln}.
We will discuss in more detail in section \ref{sec:Goff}  the reasons to assume the diagonal elements to be zero. 

\paragraph{Complex conjugation.}
All previously discussed properties of the $\bP$-functions remain unchanged under the rescaling
\begin{equation}
    \bP_1\rightarrow \lambda \bP_1,\ \bP_2\rightarrow \lambda^{-1} \bP_2\quad \bP_{\dot{1}}\rightarrow \dot{\lambda} \bP_{\dot{1}},\ \bP_{\dot{2}}\rightarrow \dot{\lambda}^{-1} \bP_{\dot{2}}
\end{equation}
where $\lambda$ and $\dot{\lambda}$ are some numbers. One can use this rescaling symmetry to choose $A_1=A_{\dot 1}=1$, so that 
that $\bP_{1},\;\bP_{\dot 1}$ are purely real and $\bP_{2},\;\bP_{\dot 2}$ are purely imaginary\footnote{The assumption of these reality properties follows from the expectation that in a unitary QFT complex conjugation should be a symmetry of the underlying model. Furthermore, this is known to be the case in the other incarnations of the QSC.}. This symmetry also translates into $\bQ$ as again we can obtain the LHPA $\bQ^\uparrow$ from $\bQ^\downarrow$ using complex conjugation. We fix the normalisation by taking $B_{ 1}$ and $B_{\dot 1}$ to be real which then implies that $B_{2}$ and $B_{\dot 2}$ are purely imaginary. Then we have
\beqa
\bQ_k^{\uparrow}(u)=(-1)^{k+1}\bar\bQ_k^{\downarrow}(u)\;\;,\;\;
\bQ_{\dot k}^{\uparrow}(u)=(-1)^{{\dot k}+1}\bar\bQ_{\dot k}^{\downarrow}(u)
\eeqa
where $\bar\bQ$ denotes the complex conjugate of $\bQ$. This implies, in combination with the gluing condition, that
\beq
\bQ_k^\downarrow(u+i0)=G_k^{\;\;\dot k}(-1)^{k+1}\bar\bQ_{\dot k}^\downarrow(u-i0)\;\;,\;\;
\bQ_{\dot k}^\downarrow(u+i0)=G_{\dot k}^{\;\;k}(-1)^{\dot k+1}\bar\bQ_{ k}^\downarrow(u-i0)\,.
\eeq
Combining the two relations we get
\beq
(-1)^{\dot k+k}G_k^{\;\;\dot k}
\bar G_{\dot k}^{\;\;l}=\delta_k^l\;.
\eeq
In combination with \eq{Galpha} this implies that $\alpha$ is real.
Finally, for $L=2$ we get the following relations
\beq\la{Nglu}
N_{k}^{\;\;\dot{l}}\equiv G_k^{\;\;\dot l}e^{-i\pi\hat M_{\dot l}}=G_{\dot k}^{\;\;l}e^{-i\pi\hat M_{l}}
=\left(
    \begin{array}{cc}
        0 & -i\, \alpha\,e^{\frac{i\pi\gamma}{2}} \\
        i\, \alpha^{-1}\,e^{-\frac{i\pi\gamma}{2}} & 0
    \end{array}
    \right)_{kl}\;.
\eeq

\paragraph{Finding $\mu^{R}$.}
In the previous paragraphs we described how to glue $\bQ$ using the gluing matrix $N_{k}{}^{\dot{l}}$. At various stages it will also be important to consider the analytic continuation of $\bP$, not only $\bQ$. To treat this we need the matrix $\mu^{R}$ appearing in \eqref{Pgamma}. To find this matrix we can rewrite the gluing condition for $\bQ$ as analytic continuation: $\bQ^{\gamma}_k = (\omega^{R})_{k}{}^{\dot{l}}\bQ_{\dot{l}}$ with $(\omega^{R})_{k}{}^{\dot{l}} = N_{k}{}^{\dot{m}}(\Omega^{R})_{\dot{m}}{}^{\dot{l}}$ and $(\Omega^{R})_{\dot{k}}{}^{\dot{l}}\bQ_{\dot{l}}(u) = \bQ_{\dot{k}}(-u)$. Then after transforming $\bQ$ into $\bP$ we will be able to read off $\mu^{R}$.

We first introduce the Q-functions $Q_{a|k}$, with $\det Q_{a|k} = 1$, and $Q^{a|k}=\epsilon^{ab}\epsilon^{kl}Q_{b|l}$ which allows us to rotate between $\bQ$ and $\bP$ according to 
\begin{align}\label{eq:QQaiP}
    &\bQ_{k} = \bP^{a}Q^{\pm}_{a|k}\,,
    &
    &\bP_{a} = Q^{\pm}_{a|k}\bQ^{k}\,.
\end{align}
The functions $Q_{a|k}$ also satisfy the fermionic QQ-relation
\begin{align}
    Q^+_{a|k} - Q^-_{a|k} = \bP_{a} \bQ_{k} ,
\end{align}
or equivalently after using \eqref{eq:QQaiP}
\begin{equation}\label{Qairelation}
    Q_{a|k}^+ -Q_{a|k}^-= -\varepsilon^{bc}\bP_a\bP_c Q^\pm_{b|k}\,.
\end{equation}
To deal with parity transformations on $\bQ_k$ we need a matrix $(\Omega^{R})_{k}{}^{l}$ such that
\begin{equation}\label{eq:QParity}
    \bQ_k(-u) = (\Omega^{R})_{k}{}^{l}\bQ_{l}(u)\,.
\end{equation}
This matrix can be constructed using \eqref{eq:QQaiP} and the parity property of $\bP_{a}$, see \eqref{eq:ParityP}, giving
\begin{equation}
    (\Omega^{R})_{k}{}^{l} = -Q_{b|k}\left(-u+\frac{\ii}{2}\right)\,\parityg_a^{\;\;b}\,Q^{a|l}\left(u+\frac{\ii}{2}\right)\,.
\end{equation}
Clearly it is impossible to fix $\Omega^R$ uniquely from \eqref{eq:QParity}, we have picked the current form by demanding that $\Omega^{R}$ implements parity and have no cut on the real axis, thus decorating it with the superscript $R$. Having obtained $\Omega^{R}$ we can use $Q_{a|k}^+$, which also does not have a cut on the real axis, to deduce the $\bP\mu$ system. This gives $\mu^{R}$ explicitly as
\begin{equation}\label{eq:muR}
    \mu^{R\;\dot{b}}_{a}(u) = -Q_{a|k}\left(u+\frac{\ii}{2}\right)\epsilon^{kl}N_{l}{}^{\dot{l}}Q_{\dot{c}|\dot{l}}\left(-u+\frac{\ii}{2}\right)\epsilon^{\dot{c}\dot{d}}\parityg_{\dot{d}}^{\;\;\dot{b}}\,.
\end{equation}

\paragraph{Summary of the problem.}
Let us summarise how to solve for the parity-symmetric states in the $\sl({2})$ sector. In words the problem can be formulated quite simply. We need to find four $\bP$ functions ($\bP_a$ and $\bP_{\dot a}$), which have one single cut $[-2g,2g]$ and satisfy the large $u$ asymptotics \eq{ppow}, such that the $\bQ$ functions derived from $\bP$'s via \eq{bax} with the asymptotics \eq{qpow} satisfy the gluing condition
\beq\label{gluingcondition}
\bQ_k(u+i0)=N_k^{\;\;\dot l}\bQ_{\dot l}(-u+i0)
\eeq
with the matrix $N$ defined in \eq{Nglu}.

In order to solve the above problem analytically at weak coupling and numerically at finite $g$, we will have to introduce some new tools in the next section.

\subsection{Off-diagonal gluing matrix and analyticity of Q-functions}\la{sec:Goff}

We will now motivate the assumption presented earlier in this subsection that the gluing matrices $G_k^{\ \dot{n}}$ and $G_{\dot{k}}^{\ n}$ are off-diagonal. This can be justified from several angles. First, this can be argued from the classical perspective. The classical spectral curve is described by quasi-momenta $p_a^{\rm A}(x),p_a^{\rm S}(x),p_{\dot{a}}^{\rm A}(x),p_{\dot{a}}^{\rm S}(x)$ where the superscripts ${\rm A}$ and ${\rm S}$ refer to ${\rm AdS}_3$ and ${\rm S}^3$ degrees of freedom, respectively. The analytical continuation of the ${\rm AdS}_3$ quasi-momenta is particularly simple and takes the form \cite{Babichenko:2009dk}
\begin{equation}\label{classicalgluing}
    \tilde p_a^{\rm A}(x)=p_a^{\rm A}\left(\frac{1}{x}\right) = p_{\dot{a}}^{\rm A}\left(x\right),\quad a\in \{1,2\}\,.
\end{equation}
The full quantum Q-functions are related to the quasi-momenta as \cite{Cavaglia:2021eqr}
\begin{equation}\label{eq:quasi}
    \left( \bQ_1,\bQ_2,\bQ_{\dot{1}},\bQ_{\dot{2}}\right) \sim \left(e^{-\int^u p_1^{\rm A}},e^{-\int^u p_2^{\rm A}},e^{-\int^u p_{\dot{2}}^{\rm A}},e^{-\int^u p_{\dot{1}}^{\rm A}} \right).
\end{equation}
Note that for the dotted $\bQ$ our conventions for indices $\dot 1$ and $\dot 2$, dictated by the convention for the large $u$ asymptotic \eq{asm}, is swapped around w.r.t. the notations of \cite{Babichenko:2009dk} for which \eq{classicalgluing} holds.
From \eqref{eq:quasi}, in combination with \eqref{classicalgluing}, we conclude that $G$ is off-diagonal as expected.

Further support for this assumption comes from weak coupling perturbation theory, which is outlined in section \ref{sec:perturbative}. In the limit $g\rightarrow 0$, $G$ needs to be strictly lower-triangular to reproduce the expected weak-coupling behaviour. Thus, the diagonal elements of $G$ should vanish both in the strong and weak coupling regimes.

To verify our assumption of an off-diagonal $G$ we perturbed our numerical solution, to be described in section \ref{sec:numerics}, away from off-diagonal $G$. We observed experimentally that our numerical solution was stable under such a perturbation if we also required $\text{tr}\, W(\pm 2g) =2$, meaning that after a few iterations $G$ would become off-diagonal if we added the diagonal elements into the set of optimised parameters. At the same time without the condition $\text{tr}\, W(\pm 2g) =2$ we are able to find a solution even for $G$ with non-zero diagonal elements. 
From this experiment, which we repeated for different states and various values of the coupling constant, we conclude, with high precision, that at least in the vicinity of the parameter space where $G$ has small diagonal elements we get the following equivalence
\beqa\la{WGtheorem}
\text{tr}\, W(\pm 2g) =2\;\;\iff\;\;G_{\dot 1}^{\;1}=G_{\dot 2}^{\;2}=0\;.
\eeqa
So, instead of trying to justify the property of $G$ being off-diagonal, we instead argue that $\text{tr}\, W(\pm 2g) =2$ as a consequence of the maximal analyticity of Q-functions, which is a part of any QSC construction.

The condition $\text{tr} W(\pm 2g)=2$ is an immediate consequence of demanding that $\bP$ is single-valued and regular at the branch-points $\pm 2g$. To see this, simply note that $\bP^{2\gamma}(\pm 2g) = \bP(\pm 2g)$ implies that $\bP$ is an eigenvector of $W(\pm 2g)$ with eigenvalue $1$. Since $\det W=1$ it follows that $\Tr\, W(\pm 2g)=2$. Such regularity assumptions are standard in QSC, and we take them to hold also in our case.\footnote{We also verified up to NNLO that our analytic weak coupling solutions satisfy these conditions.}
It would be interesting to prove the equivalence \eq{WGtheorem} using analytic argument, rather than numerics, but so far we have not found a simple proof.

The above checks only tell us that for a suitably regular $\bP$ we find isolated solutions for which $G$ is an off-diagonal matrix. At the moment we cannot rule out that there might exist exotic separated solutions corresponding to more complicated choices of $G$, but those would  probably have a weak coupling limit inconsistent with wrapping effects vanishing.

\section{New tools}\label{sec:Ptt}
In this section we introduce new tools which will help us to overcome the difficulty of the non-quadratic branch cuts, which is a new feature of this QSC.

\subsection{Why were quadratic cuts making the problem easier?}
Before explaining how to overcome the difficulty, let us first explain why it was easier to deal with quadratic cuts in the AdS${}_5$ and AdS${}_4$ QSCs.
\begin{figure}
    \centering
    \includegraphics[scale=0.135]{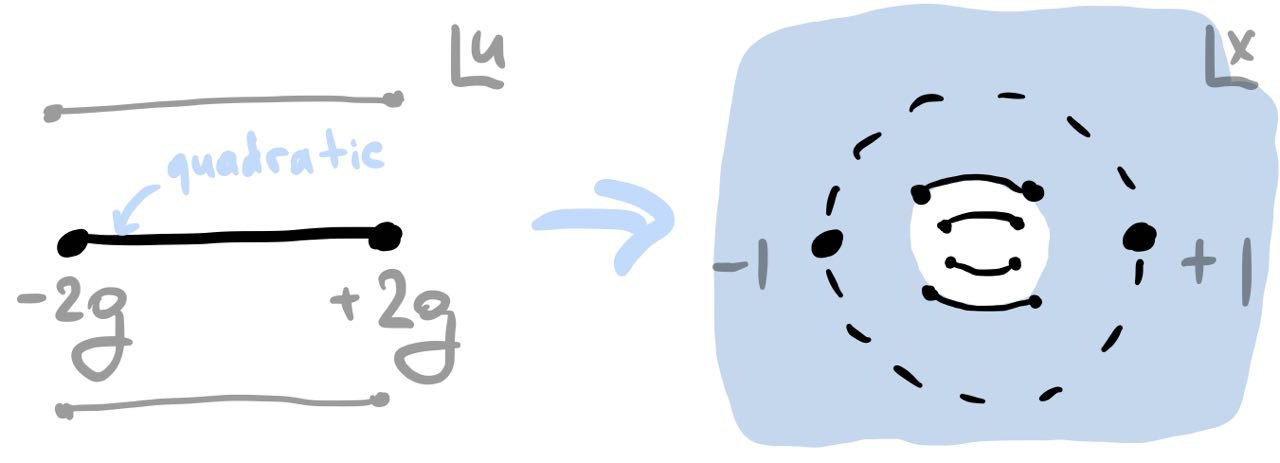}\;\;\;\;\;\;\;\;\;\;    
    \includegraphics[scale=0.135]{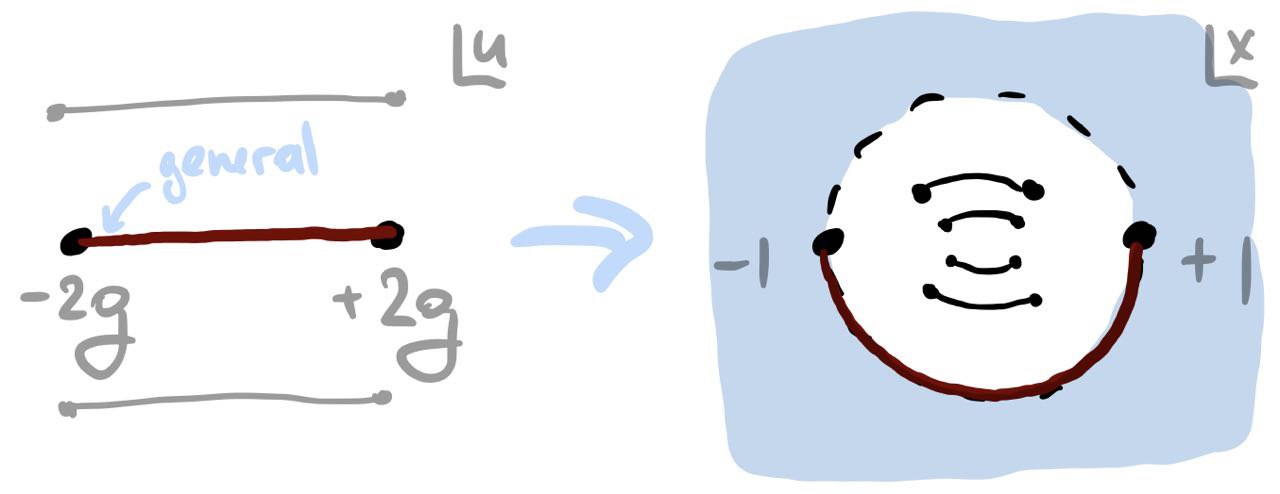}
    \caption{In the AdS${}_5$ case, the cut of the ${\bf P}$-functions is quadratic, which results in the large convergence domain (blue on the left figure) with inner radius going to zero as $g$ at weak coupling. In the current case the convergence domain stops at the remaining non-rationalized branch points at $\pm 1$ so that the inner convergence radius of the Laurent series is $1$.}
    \label{fig:cutconvergence}
\end{figure}

The $\bP$-functions in both the AdS${}_5$ and AdS${}_3$ cases have one short cut $[-2g,2g]$ on the main sheet of their Riemann surface. By introducing the Zhukovsky map $x(u)$ \eq{eqZ} this cut is blown up into a unit circle $|x|=1$. In the $x$ plane the main $u$ sheet of $\bP$ is mapped to the exterior of the unit circle and thus one can parametrize $\bP$ as a Laurent series in $x$. In the AdS${}_5$ case the square-root singularity at $u =\pm 2g$ is completely resolved in the $x$ plane and the inner convergence radius of the Laurent series is dictated by the image of the Zhukovsky cuts at $[-2g\pm i,2g \pm i]$, which appear on the second sheet of $\bP$. Since $1/x(\pm 2g\pm i)\sim \pm i g$, the inner convergence radius at weak coupling shrinks to zero, implying that at each order in $g$ there are {\it finitely} many terms in the Laurent series expansion of $\bP$, and furthermore the coefficients of the expansion in $1/x^n$ scale as $\sim g^n$.

In the case of AdS${}_3$, as we discussed repeatedly, the singularities at the branch points at $\pm 2g$ are no longer quadratic, meaning that in the $x$-plane we still have a branch cut between $\pm 1$ even at weak coupling, resulting in slower convergence of the Laurent expansion, see Figure \ref{fig:cutconvergence}. In particular, even at weak coupling we will have to include an infinite number of inverse powers of $x$. This makes it impossible to use the old strategy for solving the system perturbatively, where at each order the state was parametrised by a finite number of coefficients, which were later constrained by the gluing conditions. Similarly, this makes the old numerical approach considerably less efficient as we discuss in section~\ref{sec:numerics}.

In this section we show how to overcome this difficulty by introducing a new set of functions ${\mathbb P}$, derived from $\bP$-functions, which have a quadratic cut on the real axis. Unfortunately, the $\mathbb{P}$ functions also acquire some additional singularities as we discuss below, but, nevertheless, they will be extremely instrumental in constructing the systematic perturbative expansion and for numerics (at least for sufficiently small $g$).

\subsection{Definition and quadratic branch points}

We will now introduce a new object, which will allow us to perform the gluing efficiently. Analytic continuation of undotted $\bP$ functions $\bP_a$ takes us to dotted $\dot\bP$ functions $\bP_{\dot{a}}$  and vice versa. Hence, analytic continuation of $\bP$ functions around the same branch point twice, $\bP_a^{2\gamma}$, brings us back to a linear combination of $\bP$ functions, and this transformation can be implemented by action of a certain $2\times 2$ matrix $W$, defined in \eq{Pdoublegamma} in terms of the QSC quantities. In this subsection we will construct an object which we denote $\Ptt_a$ from $\bP$ and $W$ with the property that it is invariant under $2\gamma$, i.e. $\Ptt$ has a quadratic branch cut $[-2g,2g]$.  

For convenience let us reproduce \eqref{Pdoublegamma} here
\begin{equation}
    \left(\bP_a\right)^{2\gamma} = W_a^{\ \,b}\bP_b\,.
\end{equation}
Let us define the function $l(x)$
\begin{equation}\label{eq:lDef}
    l=\frac{i}{2\pi}\log\left(\frac{x-1}{x+1} \right)\,.
\end{equation}
Then we define the object $\Ptt$ as follows
\begin{equation}\label{eq:PttDef}
    \Ptt_a\equiv\left(W^{l} \right)_a^{\ \,b}\bP_b ,
\end{equation}
where $W^{l}$ denotes the matrix power of $W$ to power $l$. 
We recall that the matrix $W$ is regular on the real axis and the nearest singularities are the branch points at $\pm i\pm 2g$. Furthermore, at least for the states we consider, near $u\sim g$, both eigenvalues of $W$ approach $1$ at weak coupling, so the power is well defined at least for sufficiently small $g$ (we will discuss this point further below).
Let us now demonstrate that
 unlike $\bP$ where these branch points were infinite order, the branch points at $\pm 2g$ are quadratic for $\Ptt$. 
To see this, we note that under analytic continuation around the branch points $u=\pm 2g$ we have $x\rightarrow \frac{1}{x}$, which then comes back to $x$ after the second crossing of the cut. At the same time, when going around the branch point twice we cross the logarithmic cut in $l$ so that $l\rightarrow l-1$. Hence for $\Ptt$ we have
\begin{equation}
    \Ptt^{2\gamma} = W^{l-1} \bP^{2\gamma} = W^{l-1}W\bP = \Ptt\,,
\end{equation}
and the branch points in $\Ptt$ at $\pm 2g$ are quadratic!

Unlike the original $\bP$ functions it turns out that these branch points are not the only singularities of $\Ptt_a$. First of all, $W$, being built from $\mu$, has an infinite ladder of short Zhukovsky cuts in the upper and lower half plane (except on the real axis). Secondly, $W$ being raised to a non-integer power $l$ introduces other branch points. We will refer to these extra branch points as \textit{parasitic} branch points, as they are not present in the original $\bP\mu$ system and arise as a result of our construction of $\Ptt$. We will discuss them in the next subsection \ref{sec:parasitic}.

On the other hand, the fact that the branch points in $\Ptt$ at $\pm 2g$ are quadratic allows us resolve the cut in terms of the Zhukovsky variable $x(u)$ in the vicinity of the cut and express it as a Laurent series 
\begin{equation}\label{Pttlaurent}
    \Ptt_a = \displaystyle \sum_{n=-\infty}^\infty d_{a,n}x^n
\end{equation}
where $d_{a,n}$ are exponentially decreasing coefficients.
In this next subsection we discuss the analytic properties of ${\mathbb P}$ in more detail and estimate the convergence domain for the Laurent series at weak coupling. Later, we will use $\Ptt$ to solve the QSC equations numerically and analytically in sections \ref{sec:numerics} and \ref{sec:perturbative} respectively.

We also mention that there exist other combinations of $\bP$ with square-root cuts on the real axis, namely the bilinear expressions $\bP^{a}\bP_{a}^{\gamma^{n}}$. Unfortunately it turns out that imposing analytic requirements on these combinations still leaves a lot of freedom and cannot be used to fix the QSC fully, unlike ${\mathbb P}$ which will unlock for us both the analytic  perturbative regime and finite $g$ numerics. Yet, as one can expect that these objects can be utilised more efficiently in the future we have collected some results in Appendix \ref{app:MasslessBaxter}.

\subsection{Auxiliary Classical Curve}\label{sec:parasitic}
In order to analyse the analytic properties of ${\mathbb P}$, let us define the matrix power $W^l$ more explicitly. For that we use that for a general unimodular matrix $W$ with eigenvalues $\lambda,\; 1/\lambda$ we have
\beq\la{Mpow}
W^{l}=W\;\frac{m-m^{-1}}{\lambda-\lambda^{-1}}+ I\frac{m^{-1}\lambda-\lambda^{-1} m}{\lambda-\lambda^{-1}}\;\;,\;\;m=\lambda^l\;.
\eeq
The above identity is easy to verify by considering diagonal $W$ and thus is valid for all points where $W$ is diagonalisable and by continuity for all other points too.
In the above expression there are two potential sources of singularities. Firstly, the eigenvalues can be defined as a solution of a quadratic equation
\beq
T=\lambda+1/\lambda\;\;,\;\;\lambda = \frac{T+\sqrt{T^2-4}}{2},
\eeq
where $T\equiv {\rm tr}\; W$ is a regular function in the strip $-1<{\rm Im}\; u<1$. We see that $\lambda(u)$ has a branch cut each time when $T^2-4$ has an isolated zero (or in general a zero with odd multiplicity). It is convenient to introduce the ``quasi-momentum" $p(u)$ by $\lambda=e^{ip(u)}$. We have $T=2\cos p$ and by differentiating this identity we also obtain
\beq
-2p'\sin p = T'\;\;,\;\; p'=-\frac{T'}{\sqrt{4-T^2}} ,
\eeq
thus we see that $p'$ is a double-valued function 
inside the strip $-1<{\rm Im}\; u<1$. At the same time $p(u)$ can be more than double-valued due to the non-trivial monodromies of the integration contour $p=\int^u p'(w) dw$.

Using perturbative results, at least for the class of states considered here we observed\footnote{See also the discussion in section~\ref{sec:Goff}.} that $T(\pm 2g)=2$ which allows us to fix conventions so that $p(-2g)=0$. After that we can define
\beq
p=\int_{-2g}^u p'(w) dw ,
\eeq
which then defines $p$ as a function on a strip $-1<{\rm Im}\; u<1$ with additional cuts, which we call ``parasitic".

In fact the whole discussion above is reminiscent of the story of classical integrability like in the seminal paper \cite{Kazakov:2004qf}  (see also a simplified review in \cite{Gromov:2006kve}). One can speculate that this classical integrability-like structure arising here is closely linked with the massless degrees of freedom. Indeed, formally, the massless ``magnon" dispersion relation can be obtained from the massive one by taking the strong coupling limit, so it is reasonable to expect that their dynamics at finite coupling is still described by auxiliary algebraic curve type of equations.\footnote{Note that $T=\dot T$ due to the fact that $T = \text{tr}(\mu^{R} \dot \mu^{R} ) = \text{tr}(\dot \mu^{R} \mu^{R} )=\dot T$, so there is only one classical integrable system embedded into this quantum spectral curve.}

Let us consider some examples. For the ``twist-2" states we consider in this paper $T$ has the following form at the first two orders
\beq\la{Texp}
T=2+\beta g^2 (v^2-4)(v^2-1)^2+{\cal O}(g^3)\;\;,\;\;v=u/g\;.
\eeq
At this order we see several interesting features -- firstly $T(\pm 2g)=2$ as was announced before. In addition we have a double zero of $T-2$ at $u=\pm g$. We verified up to the order $g^7$ and also numerically within our precision that this zero moves at finite $g$ but remains double. This implies that $p(u)$ itself will have a proper $[-2g,2g]$ Zhukovsky cut as the double-zero sitting on the cut does not result in a singularity for $p$. At higher orders in $g$, we found that there could appear other points for which $T=2$ but they are situated far from the main cut at $v\sim 1/g$.

\begin{figure}[t]
    \centering
    \includegraphics[scale=0.2]{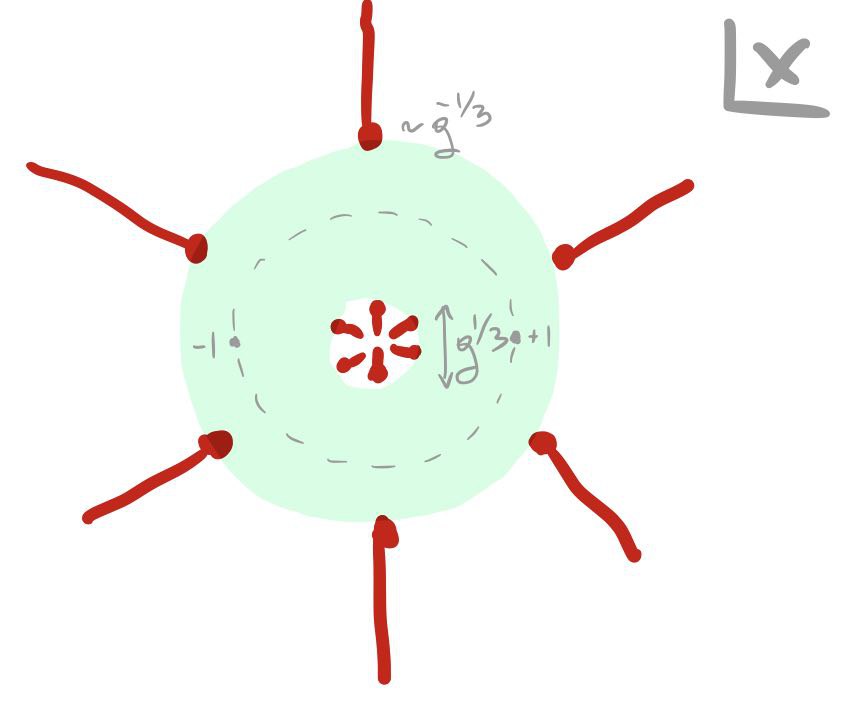}
    \caption{Parasitic cuts in the function ${\mathbb P}_a$ constrain the convergence domain (in green) to be the annulus with the inner radius scaling as $g^{1/3}$ and the outer radius scaling as $g^{-1/3}$. Importantly, the Laurent series for the function ${\mathbb P}_a$ converges fast in the vicinity of the unit circle, allowing to perform the gluing procedure for the functions living inside and outside the unit circle, such as $\bQ_k$ and $\bQ^\gamma_{\dot k}$.}
    \label{fig:my_label}
\end{figure}
Furthermore, we should also pay attention to the points where $T=-2$. It is easy to see from \eq{Texp} that there are $6$ such points situated approximately at
\beq
v^{\rm parasitic}_{a}\sim i  (-4/\beta)^{a/6} g^{-1/3}+{\cal O}(g^{1/3})\;\;,\;\;a=1,\dots,6\,.
\eeq
Based on our perturbative solution all other singularities are situated at $v\sim 1/g$. Whereas the branch cut created by $v=\pm 2$ is resolved by the Zhukovsky variable, there are the points $v^{\rm parasitic}_{a}$ which control the convergence radius for ${\mathbb P}$. Indeed, as we defined $p(-2g)=0$ and as $T=2\cos p$
and $T=-2$ at those points, we should have  $p(g v^{\rm parasitic}_{a})=\pi+2\pi n$ for some integer $n$. Rewriting the matrix power \eq{Mpow} in terms of $p(u)$ we get
\beq
W^{l}=W\;\frac{\sin(pl)}{\sin p}- I\frac{
\sin(p(l-1))}{\sin p}\;.
\eeq
We see that, whenever $p\to 0$, the matrix power does not catch the square-root singularity of $p$, as it cancels between numerator and denominator, whereas in the vicinity of $u = g v_a^{\rm parasitic}$, where $p\to \pi+2\pi n$, we have the denominator vanishing, whereas there is no general reason for the numerator to vanish anymore. Thus, $W^l$
will unavoidably have a singularities at $x\simeq v_a$ and $x\simeq 1/v_a$. From this consideration we can see that the convergence radii for the Laurent series are $\sim  g^{1/3}$ and $\sim g^{-1/3}$, which is still quite a large region at weak coupling, but is much less than the analyticity one would naively expect from the position of the Zhukovsky cuts at $\pm i\pm 2g$.

The practical implication of this is that at each order in $g$ the number of terms in the Laurent expansion for ${\mathbb P}$ is finite, but the maximal and minimal powers of $x$ would increase by roughly $3$ at each order in $g$. 

\section{Numerics}\label{sec:numerics}

In this section we will use the QSC together with the technology developed in the previous section to determine the anomalous dimension $\gamma$ of states in the $\sl(2)$ sector to high numerical precision. In order to highlight the differences between the ${\rm AdS}_3$ setup and the well-studied ${\rm AdS}_5$ case we will first review the numerical solution of the spectral problem for ${\rm AdS}_5$~\cite{Gromov:2015wca}. 

\subsection{Review of ${\rm AdS}_5$ numerical algorithm}\label{subsec:AdS5rev}

As has already been discussed in detail, the most striking distinction between the ${\rm AdS}_3$ and ${\rm AdS}_5$ integrable systems is that the branch points entering the $\bP$ functions of the latter are quadratic. In both models the $\bP$ functions on the defining sheet admit an expansion around infinity in terms of the Zhukovsky variable $x$
\begin{equation}\label{PexpansionAdS5}
    \bP_a = \displaystyle \sum_{n=-M_a}^{\infty}\frac{c_{a,n}}{x^n}\,.
\end{equation}
As we discussed in the previous section, unlike in ${\rm AdS}_3$ where this series does not converge beyond the unit circle $|x|=1$, in the ${\rm AdS}_5$ case this series converges everywhere on the first sheet and even in a finite region of the second sheet around the cut on the real axis i.e. for $|x|<1$. In this region on the second sheet the $\bP$ functions are given by simply replacing $x\rightarrow \frac{1}{x}$ in \eqref{PexpansionAdS5}.

The idea behind the ${\rm AdS}_5$ numerical algorithm is to write the expansion \eqref{PexpansionAdS5} of the $\bP$ functions up to some finite cutoff with some initial choice of the parameters $c_{a,n}$, the anomalous dimension $\gamma$ and the parameter $\alpha$ entering the gluing matrix \eqref{Nglu}. At each iteration of the algorithm these parameters get updated and converge to their true value. 
This is achieved in the following way. Using the $\bP$ functions one usually finds $Q_{a|i}$ and then $\bQ_i$ to very high numerical precision on the real axis. Similarly, one can construct the analytic continuation $\bQ_i^\gamma$ of $\bQ_i$ on the real axis and impose the gluing condition which in the ${\rm AdS}_5$ case simply takes the form (for parity invariant states)
\begin{equation}\label{ratio1}
    \frac{\bQ^\gamma_1(u)}{\bQ_3(-u)}=\alpha\,.
\end{equation}
For this to work, it is crucial that the approximation of $\bP$, given by the truncated series~\eq{PexpansionAdS5}, converges fast in the vicinity of the cut $[-2g,2g]$ or at $|x|=1$.
As in ${\rm AdS}_5$ the series is convergent for $|x|<1$, the convergence at the unit circle is exponential and the truncated series gives a very accurate approximation with relatively small number of terms in a wide range of couplings $g$.

The equation \eq{ratio1} can be reformulated as an optimisation problem
for the coefficients $c_{a,n}$. We evaluate the ratio \eqref{ratio1} at certain sampling points $u_k$, $k=1,\dots,M$, along the cut and compute its variance 
\begin{equation}
    {\bf S} = \displaystyle \sum_{k=1}^M\left|\frac{\bQ^\gamma_1(u_k)}{\bQ_3(-u_k)} -\alpha\right|\;.
\end{equation}
Then, on the true solution of the QSC this ratio is a constant so we must have ${\bf S}=0$. Hence, our goal is to minimize the function $\mathbf{S}$ by repeatedly updating the parameters $c_{a,n}$, $\gamma$ and $\alpha$, at each repetition getting closer to their true value. An efficient way to do this is via the Levenberg-Marquardt algorithm, which is described for example in \cite{Gromov:2015wca}. 
The method is quadratically convergent meaning that after each iteration the number of exact digits doubles (assuming sufficiently large cut-offs in the number of parameters).
After a few repetitions a very accurate prediction for $\gamma$ can be obtained. We refer the reader to \cite{Gromov:2015wca} for further details on the ${\rm AdS}_5$ numerical algorithm (which was also adopted in the AdS$_4$ case in \cite{Bombardelli:2018bqz}).

\subsection{Modification for the ${\rm AdS}_3$ case}

We will now describe in detail what modifications are necessary in order to treat the ${\rm AdS}_3$ case.
As was mentioned in the previous subsection, the main difference is now that while the $\bP$ functions can still be written as a convergent expansion in $x$ on the initial defining sheet~\eq{PexpansionAdS5}, this expansion is only polynomially convergent at $|x|=1$ and fails to converge for $|x|<1$. 
Hence, although we can compute $\bQ_i$ to very high precision sufficiently far from the cut on the real axis, we immediately lose precision on the cut itself as that would require evaluating the series~\eq{PexpansionAdS5} at its convergence boundary. As a result we can implement the gluing condition \eqref{gluingcondition} at the cost of a dramatic loss of precision.\footnote{Even though that is how we first found the states discussed in this paper, but with a rather unconvincing precision of a couple of digits.} 

In order to get around this we will instead implement gluing at the level of the functions $\Ptt_a$ introduced in section \ref{sec:Ptt}, which \textit{can} be written as a convergent series in $x$ near the cut. To summarise, in the ${\rm AdS}_5$ case one should follow the following sequence
\begin{equation}
    \bP\, \rightarrow\, Q_{a|i}\, \rightarrow \bQ\,  \rightarrow\, \text{gluing} ,
\end{equation}
whereas in the ${\rm AdS}_3$ case we will follow
\begin{equation}
    \bP\, \rightarrow\, Q_{a|i}\, \rightarrow \mu^R\, \rightarrow\, \Ptt\, \rightarrow \text{gluing}\,.
\end{equation}

We will now describe the procedure in detail. 

\paragraph{Parametrisation of $\bP$-functions.}
We begin by writing the $\bP$ functions as a series expansion in $x$ like in \eq{PexpansionAdS5} which in practice needs to be truncated. At the same time the upper limit $M_a$ is fixed by the quantum numbers of the state in question as in \eqref{Mnumbers}.

\paragraph{$Q_{a|k}$ at large $u$.}
Using the finite-difference equation \eqref{Qairelation} it is possible to find $Q_{a|k}$ at large $u$ as an asymptotic series expansion 
\begin{equation}\label{Qailargeu}
    Q_{a|k}\simeq u^{M_a+\hat{M}_k+1}\displaystyle\sum_{n=0}^N \frac{B_{a|k,n}}{u^n}
\end{equation}
where $N$ is some cut-off, for example $N\sim 10-20$. Plugging this expression into \eqref{Qairelation} allows to obtain the coefficients $B_{a|k,n}$ order by order numerically in terms of the $c_{a,n}$ entering the $\bP$ functions. This provides a very good approximation to $Q_{a|k}$ for large-$u$. The next step is to use this information to obtain $Q_{a|k}$ near the real axis. 

\paragraph{$Q_{a|k}$ at finite $u$.}

Next, we repeatedly use the relation \eqref{Qairelation} which can be written as 
\begin{equation}
  Q_{a|k}^-=  U_a^{\ \, b} Q_{b|k}^+,\quad U_a^{\ \, b}=\delta_a^{\ \,b} +\varepsilon^{bc} \bP_a \bP_c
\end{equation}
to obtain $Q_{a|i}(u)$ for finite-values of $u$ with high precision. Indeed, we can write
\begin{equation}\la{Qpp}
    Q_{a|k}\left(u+\frac{i}{2}\right)=\left[U(u+i)\dots U(u+i\,N')\right]_a^{\ \,b}Q_{b|k}\left(u+i\,N' +\frac{i}{2} \right)\;.
\end{equation}
For large enough $N'$ the approximation \eqref{Qailargeu} becomes valid for $Q_{b|k}\left(u+i\,N' +\frac{i}{2} \right)$ on the r.h.s. and so by repeatedly applying $U$ we obtain $Q_{a|k}$ for finite values of $u$. So far the procedure has been identical to the ${\rm AdS}_5$ case, because in order to 
use \eq{Qpp} we only need $\bP$ shifted by $i$ from the cut, where the series \eq{PexpansionAdS5} is still exponentially convergent.

\paragraph{Constructing $\mu^R$.}

Next we need to compute the quantity $\mu^R$, which controls the analytic continuation of the $\bP$ functions around their branch points on the real axis. $\mu^R$ can be expressed in terms of $Q_{a|k}$ and the gluing matrix $G$ using \eq{eq:muR}. Once we have $\mu^R$ and $\dot\mu^R$ it is also immediate to compute $W$ as their product via \eq{Wdef}. 

\paragraph{Constructing $\Ptt$ and closing the equations.}
Having constructed $W$,  we can compute $\Ptt$ simply using the definition \eq{eq:PttDef}. As this involves knowledge of $\bP$, in this way
we can only obtain $\mathbb{P}$  away from the cut, where the expansion for $\bP$ is well convergent. 

Similarly, we can compute $\Ptt^\gamma$ using
\begin{equation}\label{eq:PgammattDef}
    \Ptt^\gamma\equiv W^{l^\gamma}\mu^{R}\,\dot\bP .
\end{equation}
We recall that $W$ is regular near the cut on the real axis, whereas $l^\gamma = \frac{i}{2\pi}\log \frac{1/x-1}{1/x+1}$. 
For definiteness one can choose the log-cut in the latter expression  to connect $-1,1$
via the lower semi-circle, so that $l^\gamma$ is well defined in the interior of the unit circle. Next we recall that for ${\mathbb P}_a$ the Laurent expansion is valid everywhere in the vicinity of the unit circle (at least for small enough $g$), as discussed in Section~\ref{sec:Ptt}. This means that we should simultaneously have
\begin{equation}\label{Pttlaurent}
    \Ptt_a = \displaystyle \sum_{n=-\infty}^\infty d_{a,n} x^n\;\;{\rm and}\;\;
    \Ptt_a^\gamma = \displaystyle \sum_{n=-\infty}^\infty \frac{d_{a,n}}{x^n}\,.
\end{equation}
However this is only true when all the QSC equations are satisfied, otherwise we have to adjust the parameters $c_{a,n},\;\alpha,\;\gamma$ to bring the coefficients of the two expansions \eq{Pttlaurent} as close to each other as possible.

In practice we proceed as follows. We evaluate $\Ptt$ at some number
${\rm NP}$ of sampling points $u_k$, $k=1,\dots,{\rm NP}$ going slightly around the cut on the real axis in the $u$ plane. This is effectively achieved by choosing the $u_k$ to be on an ellipse around the cut $[-2g,2g]$ in the domain where both expansions of $\bP_a$ and ${\mathbb P}_a$ are well convergent. More specifically, we introduce $h=1+\varepsilon$ where in practice $\varepsilon \sim \frac{2}{10}$ and use the following sampling points
\begin{equation}
    u_k:=g\left(h\,e^{i\phi_k}+\frac{1}{h}e^{-i\phi_k}\right),\quad \phi_k = 2\pi \frac{k}{{\rm NP}}\,.
\end{equation}
We can numerically evaluate $\Ptt_a(u)$ at these points to high accuracy numerically using \eq{eq:PttDef}. We denote $\Ptt_{a,k}:=\Ptt_a(u_k)$. Note that those points appear on a circle in $x$ plane $x_k:=x(u_k)=he^{2\pi i\, \frac{k}{{\rm NP}}}$. Next, we notice that the coefficients $d_{a,n}$ 
can be obtained using a discrete Fourier transform $\mathcal{F}$. We can write 
\begin{equation}
    \Ptt_{a,k} = \frac{1}{{\rm NP}}\displaystyle \sum_{n=-{\rm NP}/2}^{{\rm NP}/2-1}\left(\frac{\mathcal{F}_n\left(\{\Ptt_{a,k}\}_{k=1}^{\rm NP}\right)}{h^n}\right)h^n e^{2\pi i \frac{k \, n}{{\rm NP}}}
\end{equation}
where $\mathcal{F}_n\left(\{\Ptt_{a,k}\}_{k=1}^{\rm NP}\right)$ denotes the $n$-th element of the discrete Fourier transform of the set $\{\Ptt_{a,k}\}_{k=1}^{\rm NP}$. By comparing this expression with the Laurent series \eqref{Pttlaurent} we immediately read off
$
d_{a,n}=\frac{1}{{\rm NP}h^n}\mathcal{F}_n\left(\Ptt_{a,k}\right)
$. We can repeat the same calculation with ${\mathbb P}^\gamma$, defining
$
\tilde d_{a,-n}=\frac{1}{{\rm NP}h^n}\mathcal{F}_n\left(\Ptt^\gamma_{a,k}\right)
$. When all equations are satisfied we should get
$d_{a,n}=\tilde d_{a,n}$. In practice we have to adjust our parameters $c_{a,n},\;\alpha,\;\gamma$ for this to be the case. Hence, we can again reduce the problem of fixing those parameters to an optimisation problem by introducing the vector
 $V =\left\{d_{a,n}-\tilde{d}_{a,n} \right\}_{n=-{\rm NP}/2}^{{\rm NP}/2,\; a=1,2}$ and a similar vector for dotted quantities. Then we fix the parameters, which also include the dimension $\gamma$ by minimising the quantity $\bf{S}$
\begin{equation}
    {\bf S}\equiv |V|^2+|\dot V|^2\;.
\end{equation}
where $|V|$ denotes the usual Euclidean norm of $V$. Like in the ${\rm AdS}_5$ case \cite{Gromov:2015wca} we use the Levenberg–Marquardt method to find the optimal values of the parameters.

\subsection{Results}

We will now describe the explicit results obtained using the algorithm described in the previous subsection. We consider the cases $S=2,4,6$.

In order to give an accurate portrayal of our results we will specify the precise numerical configurations we use in the algorithm in the example of $S=4$. For each $\bP_a$ function we truncate the series after $25$ $c_{a,n}$. Note that as a result of the parity properties of $\bP$ functions this effectively means we truncate the series \eqref{PexpansionAdS5} at order $\mathcal{O}\left(x^{-50}\right)$. As there are four $\bP$-functions, together with the anomalous dimension $\gamma$ and the gluing parameter $\alpha$, there are $1+1+4\times 25=102$ real parameters entering our algorithm. To accurately determine $Q_{a|i}$ at large $u$ we keep $N=26$ in the r.h.s. of \eqref{Qailargeu}. Then to shift the result to the real axis we apply \eqref{Qpp} with $N'=100$ which allows us to determine $\mu^R$ and $W$. Finally, we need to compute $\Ptt$ at some number ${\rm NP}$ of sampling points, for which we took ${\rm NP}=175$. 

By explicitly implementing this procedure we found the following values for the anomalous dimension $\gamma$, displayed in Table \ref{table:gammafors2}.  By performing a fit of our numerical data for different values of $g$ we predict the following weak-coupling expressions for $\gamma$:

\begin{align}\label{eqn:numerical1}
\begin{split}
\gamma^{S=2}= & 12.00000000000000000000 g^2+7.85770690465128972024660 g^3 \\
  &  -56.337285968283793764 g^4-129.698256905229551 g^5\\
  &+411.285855559791 g^6
    +1711.8816412 g^7-3979.81696 g^8 + \lO(g^9) 
\end{split}
\\
\nonumber
\\
\begin{split}\label{eqn:numerical2}
    \gamma^{S=4}= & 16.6666666666666666666 g^2 + 15.1576136277995558 g^3  \\
    & -90.8556401327905 g^4-226.1244406055 g^5+743.6257412g^6 \\
    & + 3144.9603 g^7  -8097.05 g^8 + \lO(g^9)
\end{split}
\\
\nonumber
\\
\begin{split}\label{eqn:numerical3}
    \gamma^{S=6}= & 19.600000000 g^2+20.962615g^3 -117.4290g^4-292.50 g^5 + \mathcal{O}(g^6)\,.
\end{split}
\end{align}

\begin{figure}[t]
\centering
  \includegraphics[width=90mm,scale=20]{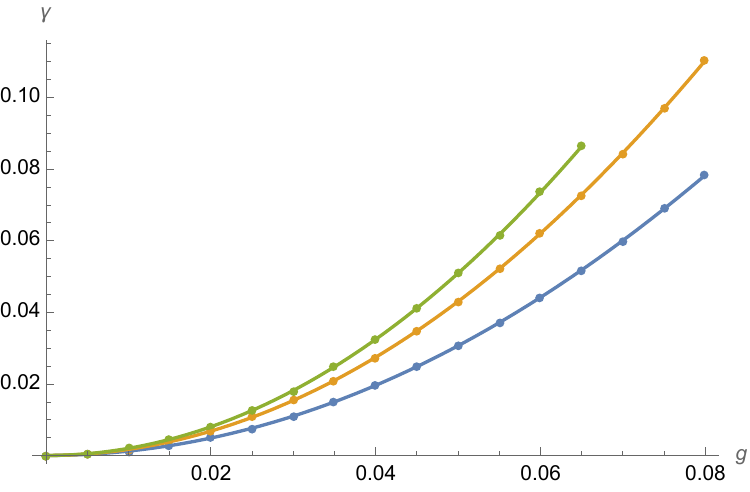}
  \caption{Numerical fit of $\gamma$ for $S=2$, $S=4$ and $S=6$. The bottom curve corresponds to $S=2$, the middle to $S=4$ and the top curve to $S=6$. }
  \label{fig:s246graphics}
\end{figure}

Finally, a plot of the anomalous dimension for $S=2,4,6$ is presented in Figure \ref{fig:s246graphics}.

\begin{table}[h]\label{table:gammafors2}
\centering 
\begin{tabular}{l l l l} 
\hline\hline 
$g$ & $\gamma^{S=2}$ & $\gamma^{S=4}$ & $\gamma^{S=6}$\\ [0.5ex] 
\hline 
0.01 & 0.00120728179220555726 &  0.001680893885675%5248864  
& 
0.00197976012314081 
\\
0.02 & 0.0048534610564808689  &   0.00677271846582 & 
0.0079880457934314
\\
0.03 & 0.010963707490747866   &  0.0153307725167  &
0.0181045829313243
\\
0.04 & 0.0195473227798079    &  0.0273845070485  &
0.032375762639963
\\
0.05 & 0.03059714740490    &  0.042936549008   & 
0.0508134671
\\
0.06 & 0.0440894161664         &  0.061962524349  & 
0.073394958
\\
0.07 & 0.059984062794             &   0.0844116699  & \\
0.08 &  0.07822544835                   &  0.11021 \\ [1ex] 
\hline 
\end{tabular}
\caption{Value of $\gamma$ for $S=2,4,6$ for fixed values of the coupling $g$. The precision drops quickly as the parasitic singularities approach the Zhukovsky cut. } 
\end{table} 

\section{Weak coupling analytic solution}\label{sec:perturbative}
In this section we explain how we solve the QSC analytically.
The procedure for AdS${}_3$ once again requires quite a substantial revision in comparison with that of the well-established AdS${}_5$ case~\cite{Marboe_2015,Marboe_2018,Marboe:2018ugv,Gromov:2015wca}, but some elements are similar.
In this section we describe the procedure we follow for solving the system analytically to obtain results for $\gamma$ for several states up to the $g^8$ order. As in general the expansion involves all integer powers of $g$ (except the linear term which appears to be zero), this would correspond to 8-loops in complexity. In practice, however, we don't know the underlying QFT description well enough to precisely count the Feynman diagrams' loops.

\subsection{The algorithm}
We will have to work with two different scales of the spectral parameter. One is when $u\sim 1$ and $x(u)\sim 1/g$ which we refer to as ``QQ-scale". Another one is when we zoom to the vicinity of the cut and keep $x(u)\sim 1$ and $u\sim g$. We will refer to this as the ``monodromy scale". We now describe the procedure in general. 

\paragraph{QQ-scale.}
In this regime one can benefit from the very fast convergent series for $\bP$-functions \eq{PexpansionAdS5}. It is better to re-expand it in terms of $u$ to obtain
\begin{equation}\label{Pexpansionu}
    \bP_a = \displaystyle \sum_{n=-M_a}^{\infty}\frac{b_{a,n}}{u^n}\,.
\end{equation}
As the Zhukovsky cut of $\bP$ shrinks to a point at small $g$, at each order in $g$ the functions $\bP_a$ are just rational functions with poles at the origin, whose order increases by roughly $1$ at each order in $g$. In other words there should be finitely many terms in the sum \eq{Pexpansionu} at each order in perturbation theory.

After that we can solve the QQ-relations in the usual way, like in AdS$_5$. The details are given in Appendix \ref{app:derivation}. In this way we obtain $Q_{a|i}^+$, $\dot{\mu}^R$ and $W$ as an expansion in $g$ where each term is a non-trivial function of $u$, expressed in terms of $\eta$-functions (which are multi-index Hurwitz zeta functions defined and reviewed in the appendix). The Q-functions at each order depend non-trivially on the unknown coefficients $b_{a,n}$ in $\bP$ \eq{Pexpansionu}.

\paragraph{Monodromy scale.}
Near the cut it is convenient to introduce $v=u/g$. For $v\sim 1$, implying $x\sim 1$, all singularities of $\dot{\mu}^R$ and $W$ move to infinity and as a result they become polynomial. At the same time $\bP_a$ become highly non-trivial.
The series \eq{Pexpansionu} contains an infinite number of coefficients even at small coupling. As the same time, we found the following double expansion
\begin{equation}\label{eq:PlExp}
    \bP_{a} = \sum_{n=0}^{\infty} l^n\bP^{(n)}_a\,,
\end{equation}
where $l=\frac{i}{2\pi}\log\(\frac{x-1}{x+1}\)$ 
and 
with $\bP^{(n)}_{a}$ containing finitely many powers of $x$ (both positive and negative) and, for the states that we consider, each suppressed by $g$ compared to the next one, i.e $\bP^{(n)}/\bP^{(n-1)} = \mathcal{O}(g)$.
In fact the expansion \eqref{eq:PlExp} follows from inverting \eqref{eq:PttDef} and expanding $W^{-l}$ for small $g$
\begin{equation}\la{Plogexp}
    \bP_{a} = (W^{-l})_{a}{}^{b}\Ptt_{b} = \sum^{\infty}_{n=0}\frac{(-l)^n}{n!}\log^n(W)_{a}{}^{b}\Ptt_b\,.
\end{equation}
As we discussed earlier, for the states we considered, the eigenvalues of $W$ go as $1+\mathcal{O}(g)$ which ensures that the series in $n$ truncates at each order in $g$.
In particular we see that the first term $\bP_a^{(0)}$, which comes without a log term, is precisely the function ${\mathbb P}$, which we introduced in section~\ref{sec:Ptt}.

Thus instead of using the infinite series \eq{PexpansionAdS5} we can parametrise $\bP_a$ in the current scale with finitely many coefficients in the polynomial $W$ and finitely many coefficients in ${\mathbb P}$ (at each order in $g$).

To summarise, we can parametrise all the relevant quantities in terms of the finitely many coefficients in ${\mathbb P}$, $W$ and $\dot\mu^R$. 
In particular we can recover $\bP_{a}$ using \eq{Plogexp}, and then $\bP_{\dot a}$ from $\bP_{\dot a} = (\mu^{R})_{\dot a}^{\;\;b}\;\bP_{b}^{\bar\gamma}$, where $\bP_{b}^{\bar\gamma}$ is easy to implement by analytically continuing $x$ to $1/x$ once we get the representation \eq{eq:PlExp}.

\paragraph{Gluing the two scales.}
So far we managed to parameterise the two scales completely independently with finitely many parameters. To close the system of equations we need to glue these two scales together. For example let us take $W$ which in the monodromy scale is represented by a series in $g$ with polynomials in $v$ as coefficients, with the polynomials growing by one degree in $v$ for each order in $g$. Let us zoom in on the largest power in $v$ at each order in $g$ where we get something of the type:
\beq
g(w_{1,6} v^6+\dots)+g^2(w_{2,7} v^7+\dots)+g^3(w_{3,8} v^8+\dots)+\dots
\eeq
where $w_{n,k}$ are some constant coefficients. We see that these leading in $v$ coefficients resum into a non-trivial function of $u= g v$
\beq
\frac{1}{g^5}
(w_{1,6} u^6+w_{2,7} u^7 +w_{3,8} u^8+\dots)+{\cal O}(g^{-4}) ,
\eeq
thus we recognise the coefficients $w_{k,k+5}$ as the expansion coefficients at the leading order in $g$ of the same function in the QQ-scale in small $u$. So in other words each term in $W$ 
in the QQ-scale contains information about infinitely many coefficients in the monodromy scale. By repeating this analysis for all $W,\;\dot \mu,\;{\bf P}$ and $\dot{\bf P}$ functions we find that we are able to gradually constrain all unknown coefficients and in particular obtain the perturbative expansion for $\gamma$.

Let us highlight that the asymptotic boundary conditions \eqref{ppow} constrains the leading parameters in \eqref{PexpansionAdS5} to depend explicitly on $\gamma$. Fixing these coefficients then also fixes $\gamma$. Finally it should be noted that the parameter $\alpha$ entering the gluing matrix $N$ is also fixed by this procedure. It enters the algorithm since $\dot\mu$ is constructed using $N$, see \eqref{eq:muR}. 

\subsection{Results}
Whereas the method above is quite general, let us highlight the results of this procedure for the two states in the $\sl(2)$ sector with $L=2$ and $S=2$ and $S=4$.
The details of the derivation for this class of operators is presented in Appendix~\ref{app:derivation}. Here we only give the result for the anomalous dimension 
\beqa
 \gamma^{S=2}&=&12 g^2+\frac{864 g^3}{35 \pi }+\left(-48-\frac{576}{7 \pi ^2}\right) g^4+\left(-\frac{405504}{875 \pi ^3}-\frac{51552}{143 \pi }\right) g^5\nonumber\\
&+&g^6\left(444-\frac{70665216}{4375
   \pi ^4}+\frac{230121984}{175175 \pi ^2}\right) \\
   &+&g^7 \left(-\frac{16896 \zeta_3}{35 \pi }-\frac{4965482496}{21875 \pi ^5}+\frac{6791453184}{875875 \pi
   ^3}+\frac{1102677696}{146965 \pi }\right)\nonumber\\
   &+&g^8 \left(-288 \zeta_3 +\frac{1898496 \zeta_3}{1225 \pi ^2}-576 \zeta_5-5844-\frac{302725824512}{109375 \pi
   ^6}\right.\nonumber\\
   &-&\left.\frac{9030729728}{25025 \pi ^4}+\frac{25695082110528}{282907625 \pi ^2}\right)\;,\nonumber\\
\gamma^{S=4}&=&
\frac{50
   g^2}{3}+\frac{1000 g^3}{21 \pi }
+\left(-\frac{1850}{27}-\frac{125000}{567 \pi ^2}\right) g^4   
   +
\left(-\frac{8800000}{5103 \pi ^3}
-\frac{43432250}{81081 \pi }\right) g^5 \nonumber\\
&+& g^6\(
\frac{1330925}{1944}-\frac{11501500000}{137781 \pi ^4}+\frac{415604390000}{45972927 \pi ^2}
\)\nonumber\\
&+& g^7 \(
-\frac{2200000 \zeta_3}{1701 \pi }-\frac{2020460000000}{1240029 \pi ^5}+\frac{48763863725000}{413756343 \pi ^3}+\frac{4915572878875}{303046029 \pi }
\)\\
&+& g^8 \(
-\frac{3250 \zeta_3}{9}+\frac{206000000 \zeta_3}{35721 \pi ^2}-\frac{10000 \zeta_5}{9}-\frac{1172878175}{69984}-\frac{923845900000000}{33480783 \pi ^6}\right. \nonumber\\
&-&\left.\frac{18365819835200000}{11171421261 \pi
   ^4}+\frac{6924759961900225}{12727933218 \pi ^2}
\)\;. \nonumber
\eeqa
These results are in perfect agreement with our numerical data \eqref{eqn:numerical1} and \eqref{eqn:numerical2}.
We notice the following curious features: firstly the expansion contains odd powers in $g$. Secondly, there are negative powers of $\pi$, which can be linked to the $\log$-type monodromies, which themselves, we believe, are related to the presence of the massless modes. In some sense perhaps one can consider $1/\pi$ as a coupling constant for the massless modes. This, however, becomes ambiguous at some order of perturbation theory as it can also contain MZVs with even arguments, which can be expressed in terms of positive powers of $\pi$. Nevertheless we observe that by sending $\pi\to\infty$ we reproduce the same results as in ${\cal N}=4$
SYM at $g^2$ and $g^4$ orders!

Curiously, the absence of the linear term is something one can in principle expect from conformal perturbation theory in symmetric product orbifolds~\cite{Apolo:2022fya}\footnote{We are grateful to A. Castro for discussing this point.}. One should be careful, however, as the coupling $g$ could have a non-trivial relation to the deformation parameter in conformal perturbation theory.

Looking at the analytic expressions above one can try to guess the general $S$ dependence. Since some of the terms coincide with ${\cal N}=4$ SYM this can act as a source of inspiration. We computed $\gamma$ up to $g^4$ for $S=2,4,6$ and $8$ and found that the data would fit the following general $S$ guess
\beq\label{eq:GeneralGammaS}
\gamma^S=
8 g^2 S_1(S)+g^3\frac{384  S_1^2(S)}{35 \pi }+
g^4\(\gamma^{\mathcal{N}=4,S}_{(4)}-\frac{512 S_1^3(S)}{21 \pi ^2}\) + \mathcal{O}(g^5)
\eeq
where for even $S$ \cite{Kotikov_2004}
\begin{equation}
    \gamma^{\mathcal{N}=4,S}_{(4)} = -16 \bigg(S_{3}(S)+S_{-3}(S)-2 S_{-2,1}(S)+2 S_{1}(S)\left[S_{2}(S)+S_{-2}(S)\right]\bigg)\,.
\end{equation}

When obtaining the above expression we assume that at order $g^n$ the transcedentality of the Harmonic sum terms cannot exceed $n-1$. We also assume that the terms with negative powers of $\pi$ should vanish as $S^2$ at small $S$, where one can argue that the effects of the massless modes should be further suppressed. We also assume that at $g^4$ the terms without $1/\pi^2$ should coincide with ${\cal N}=4$ SYM result.  In this way we still get an over-defined set of equations for some of the coefficients.

For large $S$ the harmonic number goes as $S_1(S)\sim \log S$ and we get higher powers of $\log S$ in the asymptotics, which is  totally unexpected and may signal either new physics or that the conjectured QSC contains some additional subtleties, which we are yet to understand. We notice, however, that those terms are accompanied with inverse powers of $\pi$ which is a signature of the massless modes. This phenomenon clearly deserves further more detailed exploration.

The strong coupling semi-classical analysis seems to suggest that the extra logs should be suppressed at large coupling and that the scaling in large S is similar to that in AdS$_5$, this is in contrast with the case of pure NS-NS flux \cite{Loewy:2002gf}, see \cite{David:2014qta,Banerjee:2015qeq} regarding mixed-flux. Note that the presence of the $\log^n S$ terms in the perturbative expansion does not necessarily mean that the expected $\log S$ scaling is violated at finite coupling, as those terms could be produced, for example, by some expressions of the type $e^{-g/\pi \log S}$ \footnote{An intriguing possibility is that expressions of the form $e^{-g\log S}$ might be L\"uscher-type corrections from very light particles, $m\sim g$. We thank B. Basso for bringing this to our attention.}.  In would be interesting to study the large $S$ behaviour numerically at finite $g$ and also analytically. 
At the same time, notice that the general argument about the large-spin $\log{S}$ scaling in CFTs~\cite{Alday:2007mf} is only valid in $D>2$ CFTs. We reserve this for further explorations.

\section{Discussion and Outlook}\label{sec:discussion}

Using the recently conjectured~\cite{Ekhammar:2021pys,Cavaglia:2021eqr} QSC equations for AdS$_3$/CFT$_2$ with pure Ramond-Ramond flux, we produced the first predictions for non-protected conformal data in this duality. As an important test of our conjecture we found that solutions do exist (at least for small enough coupling $g$) and furthermore those solutions are isolated (i.e. do not contain any continuous parameters for fixed coupling). This property is something highly non-trivial but of course expected from a valid construction where each solution is expected to correspond to a particular state in a CFT. We should point out, that in addition to the initial conjecture~\cite{Ekhammar:2021pys,Cavaglia:2021eqr} we imposed an extra analyticity property on the Q-functions, namely their single-valuedness at the branch points, which translates to the off-diagonality property of the gluing matrix as we discuss in section~\ref{sec:Goff}.

Our predictions include numerical high precision data as well as perturbative data to high order in the effective coupling $g$. The 
methods we developed in this paper for solving this new type of QSC are general. With the new methods one can expect to generate predictions for many states in the future. 
If the proposal for the QSC turns out to be correct, this opens very exciting possibilities for exploring the physics of this duality, and solving the longstanding mystery of the definition of the dual CFT. One could use this data to test conjectures such as the ones in \cite{Maldacena:1997re,Seiberg:1999xz,Larsen:1999uk,Pakman:2009mi,OhlssonSax:2014jtq}, and also use it as a starting point to set up a bootstrability \cite{Cavaglia:2021bnz,Cavaglia:2022qpg} program for this theory. 
Purely integrability-based methods to compute planar structure constants and correlation functions could also be on the horizon with the approaches already discussed in the Introduction \cite{Basso:2015zoa,Eden:2021xhe,Fabri:2022aup}. 

Even though our results give hope that some of these important questions could be within reach, they also raise new questions and reveal additional mysteries.
Let us close the paper by making additional comments on the form of the results, on the present evidence for the correctness of the QSC, and on future perspectives:

\begin{itemize}
    \item The pertubative expansion reveals new features. Firstly the result contains odd powers in the coupling $g$. Secondly, in addition to the multiple-zeta-values (MZVs) we also obtain inverse powers of $\pi$ in the result, which could be linked to the massless modes in our calculation. The result for the spectrum is very non-trivial and much more complicated than the pure NSNS-case.
    
    \item For both perturbative and numerical results we had to develop radical new methods to deal with non-quadratic cuts. Even though the solution of the QSC is considerably harder, the good news is that it is still solvable with the new methods we developed. We found that when solving the Riemann-Hilbert problem near the branch-cut we obtain an auxiliary structure reminiscent with classical integrability. We speculate that this classical integrability describes the massless sector. At the same time it is also responsible for the so-called \emph{parasitic} branch points which appear in the auxiliary functions we use in our algorithm. At the moment they prevent us from using the high-precision algorithm all the way to strong coupling, which is the interesting region for comparing with perturbative string theory.  
    We leave further attempts to overcome this difficulty for future studies. It would also be interesting to clarify if the Hirota-like equations we find in Appendix \ref{app:MasslessBaxter} are connected to massless modes and if they could prove useful.

\item Even though we studied only a few twist-two states, we managed to make a prediction for any value of spin $S$ at low orders in perturbation theory. We observed that, unlike in ${\cal N}=4$ SYM, at large $S$ we obtain multiple $\log S$ contributions. Those additional logs come with inverse powers of $\pi$, which hints that the massless sector could be responsible for this unusual behaviour. We note, however, that there is no immediate contradiction with the general CFT argument we can find in the literature for 2D systems, such as~\cite{Alday:2007mf}. 

\item It would be also interesting to study the analytic continuation in the spin variable $S$ to reconstruct the whole Regge trajectory like it was possible to do in the case of ${\cal N}=4$ SYM in~\cite{Gromov:2014bva}.

    \end{itemize}

There are also some possible directions to further test the QSC conjecture:
    \begin{itemize}
\item One can consider states with large length $L$. Even though it is expected that the wrapping effects are severe and are of order $1/L$, one should be able to match, at least approximately, with the ABA predictions.

\item At the moment, we do not have strong evidence that we capture massless states yet, even though it is expected that the QSC should capture those excitations too. One should be able to study states which contain massless modes in the future with the methods we developed in this paper. It would be an important test of the QSC  that these states are also present.

\item Another possible test would be to make a connection with the recently proposed set of integral TBA equations~\cite{Frolov:2021bwp}. 
As we can construct all Q-functions both numerically and analytically one can in principle get T- and Y-functions as a simple combination of these, plug them into the TBA equations and check that they are satisfied.
\end{itemize}
 
One can try to take the QSC and the methods we developed beyond the current model:
 \begin{itemize}
     \item Generalising to the NS-NS case where we have a lot of non-trivial data available to compare with. In particular, the $k=1$ setting \cite{Eberhardt:2018ouy,Eberhardt:2019ywk} could be a good playground for understanding correlators and non-planar effects with the QSC. 
     
     \item Ultimately, one should hope to extend the current formalism to the most  general mixed R-NS flux case. Integrability in this case was studied in \cite{Cagnazzo_2012} and the S-matrix found in \cite{Hoare:2013pma,Hoare:2013ida,Lloyd:2014bsa}.
     
Finally, more complicated questions include:
     
     \item 
     What changes occur in sectors with nonzero winding/momentum? What happens if one replaces ${\rm T}^4$ with $K_3$? What is the function $g(\lambda)$ and how does one fix it?
     
     \item ${\rm AdS}_3\times {\rm S}^3\times {\rm S}^3\times {\rm S}^1$. In this case the symmetry is the complicated $(d(2,1;\alpha) )^2$, and the Q-system for such an algebra is currently not known. At $\alpha = -\frac{1}{2}$, each copy of the algebra should reduce to $\mathfrak{osp}(4|2)$, where the Q-system should be a low-rank version of the one of ABJM theory written in \cite{Bombardelli:2017vhk}. This could make an interesting starting point.  
     
 \end{itemize}

We hope to return to some of these questions in the future, but it is fair to say there are still a lot of new things to uncover even within the simplest ${\rm AdS}_3\times {\rm S}^3\times {\rm T}^4$ RR-flux system.

\section*{Acknowledgements}
We thank G. Arutyunov, A. Banerjee, B. Basso, A. Belin, A. Bissi, A. Castro, M. Gaberdiel, R. Gopakumar, J. Julius, V. Kazakov, P. Kravchuk, D. le Plat, F. Levkovich-Maslyuk, S. Majumder, J. Maldacena, A. Manenti, D. Martelli, M. Preti,  A. Pribytok, N. Primi, A. Sfondrini, N. Sokolova, R. Tateo, A. Tseytlin and M. Wilhelm for discussions, and especially B.  Stefa\'nski jr, A. Torrielli and D. Volin for discussions and ongoing collaborations.  

We thank the Simons Foundation for the Varna Workshop organized by the International Center for Mathematical Sciences in Sofia, and the Kavli Institute For Theoretical Physics, Santa Barbara for hosting the program Integrability in String, Field, and Condensed Matter Theory, for warm hospitality and creating an excellent scientific environment in the final stages of this work. 

AC was partially supported by the INFN
project SFT and the EU network GATIS+, and was funded by the ERC project EXACTC during part of this work.

SE was supported by the Knut and Alice Wallenberg Foundation under grant “Exact
Results in Gauge and String Theories” Dnr KAW 2015.0083.

N.G. and P.R were supported by the European Research Council (ERC) under the European Union’s Horizon
2020 research and innovation programme – 60 – (grant agreement No. 865075) EXACTC. P.R is grateful to DESY Hamburg for warm hospitality
during the final stages of this work. 

\appendix
\section{Square-root functions and associated functional equations}\label{app:MasslessBaxter}

While $\bP_{a}$ and $\bP_{\dot{a}}$ do not have a square-root cut on the real axis, the bilinear combinations $\bP^{a}\bP^{\gamma}_{a}$ and $\bP^{\dot{a}}\bP^{\gamma}_{\dot{a}}$ do. To be precise these expressions are interchanged under $\gamma$,
\begin{equation}
    (\bP^{a}\bP_{a}^{\gamma})^{\gamma} = (\mu^{R})^{a}{}_{\dot{b}}(\mu^{R})_{a}{}^{\dot{c}}\bP^{\dot{b}} \bP^{\gamma}_{\dot{c}} = \bP^{\dot{a}}\bP^{\gamma}_{\dot{a}}\,.
\end{equation}
It is possible to generalise these expressions to a family of functions with the same square-root property
\begin{align}
    &F_{n} = \bP^{a}\bP_{a}^{\gamma^{n}}\,,
    &
    &\dot{F}_n = \bP^{\dot{a}}\bP_{\dot{a}}^{\gamma^{n}}\,,
\end{align}
where $F_n$ and $\dot{F}_n$ get interchanged under analytic continuation
\begin{align}
    &F_n^{\gamma} = \dot{F}_{n}\,,
    &
    &\dot{F}_n^{\gamma} = F_{n}\,.
\end{align}
Due to $\bP^{a}=-\epsilon^{ab}\bP_{b}$ we can rewrite $F_n$ and $\dot{F}_n$ in a determinant form. Using this form the following Hirota bilinear equation follows
\begin{equation}
    F^{\gamma}_n F_n = F_1 F_1^{\gamma^{n}} + F_{n+1}F_{n-1}^\gamma\,.
\end{equation}
Clearly these equations share similarities with a standard T-system with $F_n$ identified as T-functions and $\gamma$ playing the role of a discrete shift of the spectral parameter. It would be interesting to see if $F_n,\dot{F}_n$ can indeed be identified more concretely with a type of  transfer matrix. It is also possible to construct a version of Y-functions:
\begin{equation}
    Y_n = \frac{F_{n-1}^{\gamma}F_{n+1}}{F_1^{\gamma^n}F_{1}} ,
\end{equation}
which satisfy the following Y-system
\begin{equation}
    Y_n Y_n^{\gamma} = (1+Y^{\gamma}_{n-1}) (1+Y_{n+1})\,.
\end{equation}

We also mention that there exists a Baxter-like equation with shifts replaced by analytic continuation. It can be found by employing the standard trick of expanding a trivial determinant
\begin{equation}\label{eq:MasslessBaxter}
    \begin{vmatrix}
    \bP^{\gamma}_{a} & \bP_{a} & \bP^{\bar{\gamma}}_{a} \\
    \bP^{\gamma}_1 & \bP_{1} & \bP_{1}^{\bar{\gamma}} \\
    \bP^{\gamma}_2 & \bP_{2} & \bP_{2}^{\bar{\gamma}} 
    \end{vmatrix} = -\dot{F}_1 \bP_{a}^{\gamma}+\dot{F}_2 \bP_{a} - F_1 \bP^{\bar{\gamma}}_a = 0\,.
\end{equation}
This can also be written as an equation on $\bQ_k$:
\begin{equation}
    -\dot{F}_1\bQ_k^{\gamma}+\dot{F}_2 \bQ_k - F_1 \bQ_k^{\bar{\gamma}} = 0\,,
\end{equation}
which comes about from multiplying \eqref{eq:MasslessBaxter} with $\epsilon^{ab}Q^+_{b|k}$ and using $\bQ_{k} = \bP^{a}Q^+_{a|k}$ as well as the fact that $Q^+_{a|k}$ does not have a cut on the real axis.

\section{Details of the weak coupling perturbative expansion for $L=2$}\la{app:derivation}

In this appendix we give further details for the weak coupling algorithm we employed to find $\gamma$ analytically for several states with $L=2$ and $S$ even as described in section~\ref{sec:perturbative}. 

At the monodromy scale we took as an ansatz
\begin{align}\label{eq:PttAns}
    &\Ptt_{1} = \frac{1}{g}  \sum_{m=0}^{\infty}\sum_{\substack{n=-\Lambda_{1,m}^L\\\text{n even}}}^{\Lambda_{1,m}^H} g^m\aparam_{1,m,n} x^{n}\,,
    &
    &\Ptt_{2} = g^2 x a_{2,0,1} + g^2\sum_{m=0}^{\infty} \sum_{\substack{n-\Lambda_{2,m}^L\\\text{n odd}}}^{\Lambda_{2,m}^H}  g^m\aparam_{2,m,n} x^{n}\,,
\end{align}
with $d_{a,m,n}$ constants and $\Lambda_{1,n}^{H/L}$  finite numbers growing as $|\Lambda^{H/L}_{1/2,n+2}-\Lambda^{H/L}_{1/2,n}|=6$ due to the parasitic branch-cuts  \footnote{Fully explicitly we found $\Lambda_{1,n}^{L}=-2-6\[\frac{n}{2}\],\Lambda_{1,n}^{H}=-2+6\[\frac{n+1}{2}\],\Lambda_{2,n}^{L}=-3-6\[\frac{n}{2}\],\Lambda_{2,n}^{H}=-1+6\[\frac{n+1}{2}\]$}\,. The overall factors of $g$ in \eqref{eq:PttAns} is to match asymptotics \eqref{ppow}. The leading order of $\bP$ and $\dot{\bP}$ was fixed to
\begin{align}\label{eq:PLO}
    &\bP_{1} = \frac{A_1}{g^2 x^2 }+\mathcal{O}\(\frac{1}{g}\)\,,
    &
    &\bP_{2} = A_2 \, g \left(x - \frac{\gamma_{(2)}}{12 x^2}\(x+\frac{1}{x}\)\right)+\mathcal{O}(g^2)\,,
    \\
    &\bP_{\dot{1}} = \frac{A_{\dot{1}}}{g x}+\mathcal{O}(g^0)\,,
    &
    &\bP_{\dot{2}} = \frac{A_{\dot{2}}}{x}\,\(x+\frac{1}{x}\)+\mathcal{O}(g)\,.
\end{align}
At the QQ-scale we used an expansion of the $\bP$-functions according to
\begin{align}\label{eq:PsuExp}
    &\bP_{a} = \sum_{n=-M_a}^{\infty} \frac{b_{a,n}}{u^{n}}\,,
    &
    &\bP_{\dot{a}} = \sum_{n=-M_{\dot{a}}}^{\infty} \frac{b_{\dot{a},n}}{u^{n}}\,,
\end{align}
with the sums only running over powers that respects the parity of $\bP$. As the first singularity in the $u$ plane are the branch points at $\pm 2g$ the parameters scales as $b_{a,n}= \mathcal{O}(g^{n+M_a+1})$ and $b_{\dot{a},n} = \mathcal{O}(g^{n+M_{\dot{a}}})$, similarly to ${\rm AdS}_5$. The additional power in $b_{a,n}$ as compared to $b_{\dot{a},n}$ is due to the shortening $A_aA^{a} = \mathcal{O}(g^2)$. Relegating the details to section \ref{sec:FindingMu} we note that from this ansatz it is possible to compute $\mu^R,\dot{\mu}^R$ and $W$. In particular they are at each order in $g$ polynomials of finite degrees in $v\equiv x+\frac{1}{x}$. We required $W$, when expressed in the variable $v$, to be of the form
\begin{equation}
    W = 1 + \begin{pmatrix}
    \mathcal{O}(g) & \mathcal{O}(\frac{1}{g^2}) \\
    \mathcal{O}(g^4) & \mathcal{O}(g)
    \end{pmatrix}\,.
\end{equation}
This can be motivated from the scalings in \eqref{eq:PLO}. Since $\bP$s have definite parity $\log(W)$ is also constrained to respect this symmetry. From \eqref{Plogexp} and the fact that $l\rightarrow -l$ when $u\rightarrow -u$ we imposed that
\begin{align}
    \log(W)_{a}{}^{b}(-u) = -\mathbf{g}_{a}^{c}\log(W)_{c}{}^{d}(u)\mathbf{g}_{d}^{b}\;.
\end{align}

Having obtained $\dot{\mu}^R$ and $W$ we can construct $\bP_{a}$ from $\Ptt_a$ using \eqref{Plogexp} and $\dot{\bP}$ from $\bP^{\gamma}_{\dot{a}} = (\mu^{R})_{\dot{a}}{}^{b}\bP_{b}$. With $\bP_{a},\bP_{\dot{a}}$ at hand it is possible to match against \eqref{eq:PsuExp}. 

We observed that for $S=2,4,6,8$ the next order of $\bP$ took the following universal form
\begin{align}
    &\frac{\bP_{1}}{A_1} = \frac{1}{x^2 g^2}-\frac{1}{g}\frac{\gamma_{2}}{4\pi}\left(\frac{x^6-1}{x}\log(\frac{x-1}{x+1})+2x^4+\frac{2}{3}x^2+\frac{2}{5}-\frac{12}{7 x^2}\right)+\mathcal{O}(g^0)\,,
    \\
    &\frac{\bP_2}{A_2} = g(x-\frac{\gamma_{2}}{12 \, x^2}(x+\frac{1}{x}))+ \mathcal{O}(g^2)\,,
    \\
    &\frac{\bP_{\dot{1}}}{A_{\dot{1}}} = \frac{1}{x g}+\frac{\gamma_{2}}{4\pi}\left(\frac{x^6-1}{x^2} \log(\frac{x-1}{x+1})+2x^3+\frac{2}{3}x+\frac{2}{5 x}\right) +\mathcal{O}(g^1)\,,
    \\
    &\frac{\bP_{\dot{2}}}{A_{\dot{2}}} = (1+\frac{1}{x^2})\left(1+g\frac{\gamma_{2}}{4\pi}\left(\frac{x^6-1}{x}\log(\frac{x-1}{x+1})+2x^4+\frac{2}{3}x^2+\frac{2}{5}\right)\right)+\mathcal{O}(g^2)\,,
\end{align}
which we expect to hold for arbitrary $S$. These expressions are also useful as starting points for the numerical algorithm described in the main text. We give higher orders for $S=2$ in  \ref{sec:S2Explicit}.

\subsection{Finding $\dot{\mu}^R$ and $W$}\label{sec:FindingMu}

In this section we briefly describe how we found $\dot{\mu}^{R}$ and $W$ from the QQ-scale expansion \eqref{eq:PsuExp} of $\bP$. The methods used here are not new, they are described in detail in \cite{Marboe:2018ugv,Marboe_2018,Gromov:2015vua}

$Q^+_{a|k}$ can be computed in 2 steps, first we find a starting point by solving the Q-system at leading order in $g$ at large $u$. We found that the $\bP$-functions are simply monomials and fixed by their asymptotics:
\begin{align}\label{eq:PMonomial}
    &\bP_{1} = \frac{A_1}{u^2}\,,
    &
    &\bP_{2} = A_2 u\,,
    &
    &\bP_{\dot{1}} = \frac{A_{\dot{1}}}{u}\,,
    &
    &\bP_{\dot{2}} = A_{\dot{2}} \,.
\end{align}
As follows from \eqref{ppow} the leading coefficients have to satisfy
\begin{align*}
    &A_{1}A^{1} = -A_{2}A^{2}  = \frac{\ii}{2}\gamma_{2}\,g^2+\mathcal{O}(g^3) \,,
    &
    &A_{\dot{1}}A^{\dot{1}} = -A_{\dot{2}}A^{\dot{2}} = \ii (1+S)S+\mathcal{O}(g^2)\,, \\
    &B_{1}B^{1} = -B_{2}B^{2}  = \frac{\ii}{2}\gamma_{2}\,g^2+\mathcal{O}(g^3) \,,
    &
    &B_{\dot{1}}B^{\dot{1}} = -B_{\dot{2}}B^{\dot{2}} = \ii\frac{(1+S)S}{1+2S}+\mathcal{O}(g^2)\,.
\end{align*}
The fact that $A_1A^1 = -A_2 A^2 = \mathcal{O}(g^2)$ means that we have shortening. Using gauge-transformations we can, and will, set $A_1,A_2,B_1, B_2$ to be of order $g$.

Given \eqref{eq:PMonomial} we can find $\bQ_k,\bQ_{\dot{k}}$ by solving the Baxter equation \eqref{bax}. $\bQ_k$ is simply a rational function of $u$, explicitly
\begin{align}
    &\bQ_{1} = B_1\, u\,, 
    &
    &\bQ_{2} = B_2 \, \frac{1}{u^2}\,.
\end{align}
For the dotted system $\bQ_{\dot{1}}$ will be polynomial while $\bQ_{\dot{2}}$ will also contain so-called $\eta_{s_1 s_2\dots s_k}$-functions \cite{Leurent_2013} where 
\begin{align}
    \eta_{s_1,s_2,\dots s_k}(u) = \sum_{n_1>n_2>\dots n_k\geq 0}^{\infty} \frac{1}{(u+\ii n_1)^{s_1}\dots (u+\ii n_k)^{s_k}}
\end{align}
As an example, for $S=2$ we found
\begin{align}\label{eq:Q0S2}
    &\bQ_{\dot{1}} = B_{\dot{1}} u^2\,,
    &
    &\bQ_{\dot{2}} = \frac{5\, B_{\dot{2}}}{u} \left(-1+3 \ii u + 6 u^2 - 6 \ii u^3 \eta_{2}\right)\,.
\end{align}
From $\bP$ and $\bQ$ it is now possible to find $Q^+_{a|i}$ by solving $Q_{a|k}^+ - Q_{a|k}^- = \bP_{a}\bQ_{k}$. Notice, however, that for the undotted system we have $Q_{a|k}^+ - Q_{a|k}^- =0+ \mathcal{O}(g^2)$ so that $Q^\pm_{a|k}$ is a constant at leading order. The precise form can be fixed using \eqref{eq:QQaiP}
\begin{equation}
    Q_{a|k} = \begin{pmatrix}
    -\frac{2\ii A_1 B_1}{\gamma} & 0 \\ 0 & \frac{2\ii A_2 B_2}{\gamma}
    \end{pmatrix} + \mathcal{O}(g)\,,
\end{equation}
For the dotted system the equations on $Q_{a|i}$ are non-trivial but can be straightforwardly solved. For example, the result for $S=2$ following from \eqref{eq:PMonomial} and \eqref{eq:Q0S2} takes the form
\begin{equation}
    Q_{\dot{a}|\dot{k}} = \begin{pmatrix}
    -\frac{\ii A_{\dot{1}} B_{\dot{1}}}{2}(u^2-\frac{1}{12}) & -15 A_{\dot{1}}B_{\dot{2}}\left(\ii u+(u^2-\frac{1}{12})\eta^+_{2}\right) \\
    -\frac{\ii}{3}A_{\dot{2}}B_{\dot{1}}\, u \, (u^2+\frac{1}{4}) & -10 A_{\dot{2}}B_{\dot{2}}\left(\ii(u^2+\frac{1}{3})+u(u^2+\frac{1}{4})\eta^+_{2}\right)
    \end{pmatrix}
    + \mathcal{O}(g)\;.
\end{equation}

Once we have found the Q-system at leading order we can use the method of \cite{Gromov:2015vua} to find $Q_{a|i}$ to arbitrary order by solving 
\begin{equation}
    Q^+_{a|i}-Q^-_{a|i} = \bP_a \bQ_i
\end{equation}
iteratively. From this data we can then construct $\dot{\mu}^{R},\mu^{R}$ and $W$. These objects will naturally be constructed at the QQ-scale, so to find their expansion at the monodromy scale requires changing variables from $u$ to $v=\frac{u}{g}$. Conveniently, $Q^{+}_{a|k}$, since it is an object without cuts on the real axis, will expand as a polynomial in $u$ and thus also $v$ close to $\pm 2g$. 

\subsection{Details for $S=2$}\label{sec:S2Explicit}
To exemplify our general procedure we give in this subsection explicit expressions for some of the various objects that we have introduced previously in the case $S=2$. Due to increasing complexity and size of the expressions we report here only expressions to second non-trivial order.

For $\Ptt$ we found
\begin{equation}
\begin{split}
    \Ptt_1/A_1 &= \frac{1}{g^2 x^2} -\frac{9}{2 x^8}
    -\frac{2}{35 g \pi x^2}(-90+21x^2+35x^4+105x^6) \\
    &-\frac{2}{1225 \pi ^2 x^6} \left(45360 x^{10}-12075 x^8+742 x^6-68329 x^4+7350 x^2+11025\right)+\mathcal{O}(g) \\
\end{split}
\end{equation}
\begin{equation}
\begin{split}
    \Ptt_2/A_2 &= \frac{g}{x^3}(x^4-x^2-1)
    +\frac{2g^2}{5 \pi x^3}(-3-10x^2-7x^4+20x^6+15x^8)\\
    &\frac{g^3}{2 x^9}(-12 x^8-18 x^6-5 x^4+9 x^2+9)
    \\
    &+\frac{2 g^3}{25 \pi ^2 x^7} \left(855 x^{12}+735 x^{10}-264 x^8-1305 x^6-976 x^4+375 x^2+225\right)+\mathcal{O}(g^4)
\end{split}
\end{equation}
To be able to write the expressions in a relatively simple form we introduce the following notation
\begin{align}
    &\bigg[f(x)\bigg]_{n} = f(x)-\sum_{m}^{n} \frac{c_m}{x^m}\,,
    &
    &f(x)\bigg|_{x\rightarrow\infty} = \sum_{m} \frac{c_m}{x^m}\,.
\end{align}
The functions $\bP,\dot{\bP}$ are given by
\begin{equation}
\begin{split}
    \frac{\bP_{1}}{A_1} &= 
    \frac{1}{x^2 g^2}-\frac{12}{x^4 \pi^2}-\frac{18}{x^6\pi^2}-\frac{9}{2 x^8} \\
    &+\frac{3}{g \pi}\bigg[\frac{(1-x^6)}{x}\log(\frac{x-1}{x+1})\bigg]_2\\
    &+\frac{6}{35 \pi^2}\bigg[(x^2-1)\left(\frac{105}{x^7}+\frac{140}{x^5}+\frac{161}{x^3}-\frac{160}{x}-139 x-216 x^3\right)\log\frac{x-1}{x+1}\bigg]_2\\
    &-\frac{9}{2\pi^2}\bigg[ \frac{(x^6-1)^2}{x^8}\log^2\frac{x-1}{x+1} \bigg]_{2} + \mathcal{O}(g)
\end{split}
\end{equation}
\begin{equation}
\begin{split}
    \frac{\bP_{2}}{A_2} &= 
    g\left(x-\frac{1}{x}-\frac{1}{x^3}\right)-g^2\left(\frac{6}{5 x^3}+\frac{4}{x}\right)
    \\
    &+\frac{g^3}{\pi}(\frac{9}{2 x^9}+\frac{9}{2 x^7}-\frac{5}{2 x^5}-\frac{9}{x^3}-\frac{6}{x}+\frac{18}{\pi^2 x^7}+\frac{30}{\pi^2 x^5}-\frac{1952}{25 \pi^2 x^3}-\frac{522}{5 \pi^2 x}) \\
    &+\frac{3 g^2}{\pi}\bigg[\frac{(x^6-1)(x^4+x^2-1)}{x^4}\log \frac{x-1}{x+1}\bigg]_{-1} \\
   &+\frac{6}{5\pi^2}g^3\bigg[\frac{(x^2-1)}{x^8}(-15-35x^2-46x^4+5x^6+30x^8+41x^{10}+21x^{12})\log\frac{x-1}{x+1}\bigg]_{-1}\\
    &-\frac{9 g^3}{2\pi^2}\bigg[ \frac{(x^6-1)^2(x^4-x^2-1)}{x^9}\log^2\frac{x-1}{x+1} \bigg]_{-1}+ \mathcal{O}(g^4)
\end{split}
\end{equation}
\begin{equation}
\begin{split}
    \frac{\bP_{\dot{1}}}{A_{\dot{1}}} &= \frac{1}{x g}-g(\frac{3}{x^3}+\frac{9}{2x^7})-\frac{g}{\pi^2}(\frac{12}{x^3}+\frac{18}{x^5}) \\
    &+\frac{3}{\pi}\bigg[\frac{x^6-1}{x^2}\log \frac{x-1}{x+1}\bigg]_{1}
    \\
    &+g\frac{6}{5 \pi^2}\bigg[\frac{x^2-1}{x^6}(15+20x^2+41x^4+18x^6+21x^8)\log \frac{x-1}{x+1}\bigg]_{1} \\
    &-\frac{9 g}{2\pi^2}\bigg[\frac{(x^6-1)^2}{x^7}\log^2 \frac{x-1}{x+1}\bigg]_1+ \mathcal{O}(g^2)
\end{split}    
\end{equation}
\\
\begin{equation}
\begin{split}
    \frac{\bP_{\dot{2}}}{A_{\dot{2}}} &= 1+\frac{1}{x^2}+g\frac{6}{5\pi x^2}+g^2(-\frac{9}{2x^8}-\frac{9}{2x^6}-\frac{2}{x^4}-\frac{3}{x^2})+\frac{g^2}{\pi^2}(-\frac{18}{x^6}-\frac{30}{x^4}+\frac{1502}{25 x^2}) \\
    &+\frac{3g}{\pi}\bigg[\frac{(x^6-1)(x^2+1)}{x^3}\log\frac{x-1}{x+1}\bigg]_{0} \\
    &+\frac{6 g^2}{5 \pi^2}\bigg[\frac{(x^2+1)(x^2-1)}{x^7}(15+20x^2+41x^4+18x^6+21x^8)\log \frac{x-1}{x+1}\bigg]_0 \\
    &-\frac{9 g^2}{2\pi^2}\bigg[\frac{(x^4-1)(1+x^2+x^4)(x^6-1)}{x^8}\log^2\frac{x-1}{x+1}\bigg]_{0}+ \mathcal{O}(g^3)
\end{split}    
\end{equation}
They explicitly show that despite the logs $\bP$ is regular at the branch-points implying the property $\bP^{2\gamma}(\pm 2g)=\bP(2g)$.
The matrix $\mu_{\dot{a}}{}^{b}$ close to the cut is obtained as 
\begin{equation}
\begin{split}
    &(\mu^{R})_{\dot{a}}{}^{b}/\begin{pmatrix}\frac{A_{\dot{1}}}{A_1} & \frac{A_{\dot{1}}}{A_2} \\ \frac{A_{\dot{2}}}{A_{1}} & \frac{A_{\dot{2}}}{A_{2}}\end{pmatrix}\\
    &= \begin{pmatrix}
    v g+(-6\ii +\frac{324 v}{35 \pi})g^2+4v^3  & \frac{1}{g^2}+\frac{36}{5\pi g}-\frac{3\ii v}{g} \\
    g^2 v^2 + \frac{6}{35}v(-35 \ii + \frac{54}{\pi}v)g^3  & \frac{v}{g}+\frac{3 v}{5 \pi}(12- 5 \ii \pi v)
    \end{pmatrix} \\
    &+\begin{pmatrix}
    (-\frac{1104\ii}{35 \pi}-22 v+\frac{23176}{175 \pi^2})g^3 & -3 v^2-\frac{108 i v}{5 \pi }+\frac{3112}{25 \pi ^2}-10 \\
   g^4 \left(5 v^4+\frac{23176 v^2}{175 \pi ^2}-25 v^2-\frac{1104 i v}{35 \pi }-1\right)& -\frac{g v \left(25 \pi ^2 \left(2 v^2+13\right)+540 i \pi  v-3112\right)}{25 \pi ^2}
    \end{pmatrix} \\
    &+\begin{pmatrix}
        \mathcal{O}(g^4) & \mathcal{O}(g) \\
        \mathcal{O}(g^5) & \mathcal{O}(g^2)
        \end{pmatrix}
\end{split}
\end{equation}
while $W$ is computed as 
\begin{equation}
    \begin{split}
        &W_{a}{}^{b}/\begin{pmatrix}
        1 & \frac{A_1}{A_2} \\
        \frac{A_2}{A_1} & 1
        \end{pmatrix}\\
        &= \begin{pmatrix}
            1 - 6 \ii g(v-4v^3+v^5) & -\frac{6 \ii}{g^2}(2-4v^2+v^4)-\frac{48 i \left(54 v^4-279 v^2+199\right)}{35 \pi  g} \\
           6 i g^4 \left(v^6-4 v^4+v^2-2\right)+ \frac{48 i g^5 \left(72 v^6-351 v^4+226 v^2+17\right)}{35 \pi } & 1+6 i g \left(v^5-4 v^3+v\right)
            \end{pmatrix}
    \\
    &\quad \quad \quad +\begin{pmatrix}
    \mathcal{O}(g^2) & \mathcal{O}(g^0) \\
    \mathcal{O}(g^6) & \mathcal{O}(g^2)
    \end{pmatrix}\,.
    \end{split}
\end{equation}

\bibliographystyle{JHEP}
\bibliography{ref}

\providecommand{\noopsort}[1]{}\providecommand{\singleletter}[1]{#1}%

\providecommand{\href}[2]{#2}\begingroup\raggedright\begin{thebibliography}{10}

\bibitem{Maldacena:1997re}
J.~M. Maldacena, \emph{{The Large N limit of superconformal field theories and
  supergravity}}, \href{https://doi.org/10.1023/A:1026654312961}{\emph{Adv.
  Theor. Math. Phys.} {\bfseries 2} (1998) 231}
  [\href{https://arxiv.org/abs/hep-th/9711200}{{\ttfamily hep-th/9711200}}].

\bibitem{Eberhardt:2018ouy}
L.~Eberhardt, M.~R. Gaberdiel and R.~Gopakumar, \emph{{The Worldsheet Dual of
  the Symmetric Product CFT}},
  \href{https://doi.org/10.1007/JHEP04(2019)103}{\emph{JHEP} {\bfseries 04}
  (2019) 103} [\href{https://arxiv.org/abs/1812.01007}{{\ttfamily
  1812.01007}}].

\bibitem{Eberhardt:2019ywk}
L.~Eberhardt, M.~R. Gaberdiel and R.~Gopakumar, \emph{{Deriving the
  AdS$_{3}$/CFT$_{2}$ correspondence}},
  \href{https://doi.org/10.1007/JHEP02(2020)136}{\emph{JHEP} {\bfseries 02}
  (2020) 136} [\href{https://arxiv.org/abs/1911.00378}{{\ttfamily
  1911.00378}}].

\bibitem{Eberhardt:2019qcl}
L.~Eberhardt and M.~R. Gaberdiel, \emph{{String theory on AdS$_3$ and the
  symmetric orbifold of Liouville theory}},
  \href{https://doi.org/10.1016/j.nuclphysb.2019.114774}{\emph{Nucl. Phys. B}
  {\bfseries 948} (2019) 114774}
  [\href{https://arxiv.org/abs/1903.00421}{{\ttfamily 1903.00421}}].

\bibitem{Seiberg:1999xz}
N.~Seiberg and E.~Witten, \emph{{The D1 / D5 system and singular CFT}},
  \href{https://doi.org/10.1088/1126-6708/1999/04/017}{\emph{JHEP} {\bfseries
  04} (1999) 017} [\href{https://arxiv.org/abs/hep-th/9903224}{{\ttfamily
  hep-th/9903224}}].

\bibitem{Larsen:1999uk}
F.~Larsen and E.~J. Martinec, \emph{{U(1) charges and moduli in the D1 - D5
  system}}, \href{https://doi.org/10.1088/1126-6708/1999/06/019}{\emph{JHEP}
  {\bfseries 06} (1999) 019}
  [\href{https://arxiv.org/abs/hep-th/9905064}{{\ttfamily hep-th/9905064}}].

\bibitem{Pakman:2009mi}
A.~Pakman, L.~Rastelli and S.~S. Razamat, \emph{{A Spin Chain for the Symmetric
  Product CFT(2)}}, \href{https://doi.org/10.1007/JHEP05(2010)099}{\emph{JHEP}
  {\bfseries 05} (2010) 099} [\href{https://arxiv.org/abs/0912.0959}{{\ttfamily
  0912.0959}}].

\bibitem{OhlssonSax:2014jtq}
O.~Ohlsson~Sax, A.~Sfondrini and B.~Stefanski, \emph{{Integrability and the
  Conformal Field Theory of the Higgs branch}},
  \href{https://doi.org/10.1007/JHEP06(2015)103}{\emph{JHEP} {\bfseries 06}
  (2015) 103} [\href{https://arxiv.org/abs/1411.3676}{{\ttfamily 1411.3676}}].

\bibitem{Maldacena:2000hw}
J.~M. Maldacena and H.~Ooguri, \emph{{Strings in AdS(3) and SL(2,R) WZW model
  1.: The Spectrum}}, \href{https://doi.org/10.1063/1.1377273}{\emph{J. Math.
  Phys.} {\bfseries 42} (2001) 2929}
  [\href{https://arxiv.org/abs/hep-th/0001053}{{\ttfamily hep-th/0001053}}].

\bibitem{Babichenko:2009dk}
A.~Babichenko, B.~Stefanski, Jr. and K.~Zarembo, \emph{{Integrability and the
  AdS(3)/CFT(2) correspondence}},
  \href{https://doi.org/10.1007/JHEP03(2010)058}{\emph{JHEP} {\bfseries 03}
  (2010) 058} [\href{https://arxiv.org/abs/0912.1723}{{\ttfamily 0912.1723}}].

\bibitem{Sfondrini:2014via}
A.~Sfondrini, \emph{{Towards integrability for ${\rm Ad}{{{\rm S}}_{{\bf
  3}}}/{\rm CF}{{{\rm T}}_{{\bf 2}}}$}},
  \href{https://doi.org/10.1088/1751-8113/48/2/023001}{\emph{J. Phys. A}
  {\bfseries 48} (2015) 023001}
  [\href{https://arxiv.org/abs/1406.2971}{{\ttfamily 1406.2971}}].

\bibitem{OhlssonSax:2011ms}
O.~Ohlsson~Sax and B.~Stefanski, Jr., \emph{{Integrability, spin-chains and the
  AdS3/CFT2 correspondence}},
  \href{https://doi.org/10.1007/JHEP08(2011)029}{\emph{JHEP} {\bfseries 08}
  (2011) 029} [\href{https://arxiv.org/abs/1106.2558}{{\ttfamily 1106.2558}}].

\bibitem{Borsato:2014exa}
R.~Borsato, O.~Ohlsson~Sax, A.~Sfondrini and B.~Stefanski, \emph{{Towards the
  All-Loop Worldsheet S Matrix for $AdS_3\times S^3\times T^4$}},
  \href{https://doi.org/10.1103/PhysRevLett.113.131601}{\emph{Phys. Rev. Lett.}
  {\bfseries 113} (2014) 131601}
  [\href{https://arxiv.org/abs/1403.4543}{{\ttfamily 1403.4543}}].

\bibitem{Borsato:2014hja}
R.~Borsato, O.~Ohlsson~Sax, A.~Sfondrini and B.~Stefanski, \emph{{The complete
  AdS$_{3} \times$ S$^3 \times$ T$^4$ worldsheet S matrix}},
  \href{https://doi.org/10.1007/JHEP10(2014)066}{\emph{JHEP} {\bfseries 10}
  (2014) 066} [\href{https://arxiv.org/abs/1406.0453}{{\ttfamily 1406.0453}}].

\bibitem{Borsato:2016kbm}
R.~Borsato, O.~Ohlsson~Sax, A.~Sfondrini and B.~Stefa\'nski, \emph{{On the
  spectrum of AdS$_3$ \texttimes{} S$^3$ \texttimes{} T$^4$ strings with
  Ramond\textendash{}Ramond flux}},
  \href{https://doi.org/10.1088/1751-8113/49/41/41LT03}{\emph{J. Phys. A}
  {\bfseries 49} (2016) 41LT03}
  [\href{https://arxiv.org/abs/1605.00518}{{\ttfamily 1605.00518}}].

\bibitem{Borsato:2016xns}
R.~Borsato, O.~Ohlsson~Sax, A.~Sfondrini, B.~Stefa\'nski and A.~Torrielli,
  \emph{{On the dressing factors, Bethe equations and Yangian symmetry of
  strings on AdS$_3 \times$ S$^3 \times$ T$^4$}},
  \href{https://doi.org/10.1088/1751-8121/50/2/024004}{\emph{J. Phys. A}
  {\bfseries 50} (2017) 024004}
  [\href{https://arxiv.org/abs/1607.00914}{{\ttfamily 1607.00914}}].

\bibitem{Ekhammar:2021pys}
S.~Ekhammar and D.~Volin, \emph{{Monodromy bootstrap for $SU(2|2)$ quantum
  spectral curves: from Hubbard model to AdS$_{3}$/CFT$_{2}$}},
  \href{https://doi.org/10.1007/JHEP03(2022)192}{\emph{JHEP} {\bfseries 03}
  (2022) 192} [\href{https://arxiv.org/abs/2109.06164}{{\ttfamily
  2109.06164}}].

\bibitem{Cavaglia:2021eqr}
A.~Cavagli\`a, N.~Gromov, B.~Stefa\'nski, jr. and A.~Torrielli, \emph{{Quantum
  Spectral Curve for AdS$_{3}$/CFT$_{2}$: a proposal}},
  \href{https://doi.org/10.1007/JHEP12(2021)048}{\emph{JHEP} {\bfseries 12}
  (2021) 048} [\href{https://arxiv.org/abs/2109.05500}{{\ttfamily
  2109.05500}}].

\bibitem{Beisert:2010jr}
N.~Beisert et~al., \emph{{Review of AdS/CFT Integrability: An Overview}},
  \href{https://doi.org/10.1007/s11005-011-0529-2}{\emph{Lett. Math. Phys.}
  {\bfseries 99} (2012) 3} [\href{https://arxiv.org/abs/1012.3982}{{\ttfamily
  1012.3982}}].

\bibitem{Gromov:2017blm}
N.~Gromov, \emph{{Introduction to the Spectrum of $N=4$ SYM and the Quantum
  Spectral Curve}},  \href{https://arxiv.org/abs/1708.03648}{{\ttfamily
  1708.03648}}.

\bibitem{Abbott:2015pps}
M.~C. Abbott and I.~Aniceto, \emph{{Massless L\"uscher terms and the
  limitations of the AdS$_3$ asymptotic Bethe ansatz}},
  \href{https://doi.org/10.1103/PhysRevD.93.106006}{\emph{Phys. Rev. D}
  {\bfseries 93} (2016) 106006}
  [\href{https://arxiv.org/abs/1512.08761}{{\ttfamily 1512.08761}}].

\bibitem{Abbott:2020jaa}
M.~C. Abbott and I.~Aniceto, \emph{{Integrable field theories with an
  interacting massless sector}},
  \href{https://doi.org/10.1103/PhysRevD.103.086017}{\emph{Phys. Rev. D}
  {\bfseries 103} (2021) 086017}
  [\href{https://arxiv.org/abs/2002.12060}{{\ttfamily 2002.12060}}].

\bibitem{Gromov:2013pga}
N.~Gromov, V.~Kazakov, S.~Leurent and D.~Volin, \emph{{Quantum Spectral Curve
  for Planar $\mathcal{N} = 4$ Super-Yang-Mills Theory}},
  \href{https://doi.org/10.1103/PhysRevLett.112.011602}{\emph{Phys. Rev. Lett.}
  {\bfseries 112} (2014) 011602}
  [\href{https://arxiv.org/abs/1305.1939}{{\ttfamily 1305.1939}}].

\bibitem{Gromov:2014caa}
N.~Gromov, V.~Kazakov, S.~Leurent and D.~Volin, \emph{{Quantum spectral curve
  for arbitrary state/operator in AdS$_{5}$/CFT$_{4}$}},
  \href{https://doi.org/10.1007/JHEP09(2015)187}{\emph{JHEP} {\bfseries 09}
  (2015) 187} [\href{https://arxiv.org/abs/1405.4857}{{\ttfamily 1405.4857}}].

\bibitem{Cavaglia:2014exa}
A.~Cavagli\`a, D.~Fioravanti, N.~Gromov and R.~Tateo, \emph{{Quantum Spectral
  Curve of the $\mathcal N=$ 6 Supersymmetric Chern-Simons Theory}},
  \href{https://doi.org/10.1103/PhysRevLett.113.021601}{\emph{Phys. Rev. Lett.}
  {\bfseries 113} (2014) 021601}
  [\href{https://arxiv.org/abs/1403.1859}{{\ttfamily 1403.1859}}].

\bibitem{Bombardelli:2017vhk}
D.~Bombardelli, A.~Cavagli\`a, D.~Fioravanti, N.~Gromov and R.~Tateo,
  \emph{{The full Quantum Spectral Curve for $AdS_4/CFT_3$}},
  \href{https://doi.org/10.1007/JHEP09(2017)140}{\emph{JHEP} {\bfseries 09}
  (2017) 140} [\href{https://arxiv.org/abs/1701.00473}{{\ttfamily
  1701.00473}}].

\bibitem{Gromov:2015wca}
N.~Gromov, F.~Levkovich-Maslyuk and G.~Sizov, \emph{{Quantum Spectral Curve and
  the Numerical Solution of the Spectral Problem in AdS5/CFT4}},
  \href{https://doi.org/10.1007/JHEP06(2016)036}{\emph{JHEP} {\bfseries 06}
  (2016) 036} [\href{https://arxiv.org/abs/1504.06640}{{\ttfamily
  1504.06640}}].

\bibitem{Marboe:2014gma}
C.~Marboe and D.~Volin, \emph{{Quantum spectral curve as a tool for a
  perturbative quantum field theory}},
  \href{https://doi.org/10.1016/j.nuclphysb.2015.08.021}{\emph{Nucl. Phys. B}
  {\bfseries 899} (2015) 810}
  [\href{https://arxiv.org/abs/1411.4758}{{\ttfamily 1411.4758}}].

\bibitem{Gromov:2015vua}
N.~Gromov, F.~Levkovich-Maslyuk and G.~Sizov, \emph{{Pomeron Eigenvalue at
  Three Loops in $\mathcal N=$ 4 Supersymmetric Yang-Mills Theory}},
  \href{https://doi.org/10.1103/PhysRevLett.115.251601}{\emph{Phys. Rev. Lett.}
  {\bfseries 115} (2015) 251601}
  [\href{https://arxiv.org/abs/1507.04010}{{\ttfamily 1507.04010}}].

\bibitem{Cavaglia:2018lxi}
A.~Cavagli\`a, N.~Gromov and F.~Levkovich-Maslyuk, \emph{{Quantum spectral
  curve and structure constants in $ \mathcal{N}=4 $ SYM: cusps in the ladder
  limit}}, \href{https://doi.org/10.1007/JHEP10(2018)060}{\emph{JHEP}
  {\bfseries 10} (2018) 060}
  [\href{https://arxiv.org/abs/1802.04237}{{\ttfamily 1802.04237}}].

\bibitem{Giombi:2018hsx}
S.~Giombi and S.~Komatsu, \emph{{More Exact Results in the Wilson Loop Defect
  CFT: Bulk-Defect OPE, Nonplanar Corrections and Quantum Spectral Curve}},
  \href{https://doi.org/10.1088/1751-8121/ab046c}{\emph{J. Phys. A} {\bfseries
  52} (2019) 125401} [\href{https://arxiv.org/abs/1811.02369}{{\ttfamily
  1811.02369}}].

\bibitem{Cavaglia:2021mft}
A.~Cavagli\`a, N.~Gromov and F.~Levkovich-Maslyuk, \emph{{Separation of
  variables in AdS/CFT: functional approach for the fishnet CFT}},
  \href{https://doi.org/10.1007/JHEP06(2021)131}{\emph{JHEP} {\bfseries 06}
  (2021) 131} [\href{https://arxiv.org/abs/2103.15800}{{\ttfamily
  2103.15800}}].

\bibitem{Bercini:2022jxo}
C.~Bercini, A.~Homrich and P.~Vieira, \emph{{Structure Constants in
  $\mathcal{N} = 4$ SYM and Separation of Variables}},
  \href{https://arxiv.org/abs/2210.04923}{{\ttfamily 2210.04923}}.

\bibitem{Basso:2022nny}
B.~Basso, A.~Georgoudis and A.~K. Sueiro, \emph{{Structure constants of short
  operators in planar $\mathcal{N}=4$ SYM theory}},
  \href{https://arxiv.org/abs/2207.01315}{{\ttfamily 2207.01315}}.

\bibitem{Cavaglia:2021bnz}
A.~Cavagli\`a, N.~Gromov, J.~Julius and M.~Preti, \emph{{Integrability and
  conformal bootstrap: One dimensional defect conformal field theory}},
  \href{https://doi.org/10.1103/PhysRevD.105.L021902}{\emph{Phys. Rev. D}
  {\bfseries 105} (2022) L021902}
  [\href{https://arxiv.org/abs/2107.08510}{{\ttfamily 2107.08510}}].

\bibitem{Cavaglia:2022qpg}
A.~Cavagli\`a, N.~Gromov, J.~Julius and M.~Preti, \emph{{Bootstrability in
  defect CFT: integrated correlators and sharper bounds}},
  \href{https://doi.org/10.1007/JHEP05(2022)164}{\emph{JHEP} {\bfseries 05}
  (2022) 164} [\href{https://arxiv.org/abs/2203.09556}{{\ttfamily
  2203.09556}}].

\bibitem{Caron-Huot:2022sdy}
S.~Caron-Huot, F.~Coronado, A.-K. Trinh and Z.~Zahraee, \emph{{Bootstrapping
  $\mathcal{N}=4$ sYM correlators using integrability}},
  \href{https://arxiv.org/abs/2207.01615}{{\ttfamily 2207.01615}}.

\bibitem{Eden:2021xhe}
B.~Eden, D.~l. Plat and A.~Sfondrini, \emph{{Integrable bootstrap for
  AdS$_{3}$/CFT$_{2}$ correlation functions}},
  \href{https://doi.org/10.1007/JHEP08(2021)049}{\emph{JHEP} {\bfseries 08}
  (2021) 049} [\href{https://arxiv.org/abs/2102.08365}{{\ttfamily
  2102.08365}}].

\bibitem{Fabri:2022aup}
M.~Fabri, \emph{{Hexagonalization in $AdS_3 \times S^3 \times T^4$: Mirror
  Corrections}},  \href{https://arxiv.org/abs/2209.01959}{{\ttfamily
  2209.01959}}.

\bibitem{Basso:2015zoa}
B.~Basso, S.~Komatsu and P.~Vieira, \emph{{Structure Constants and Integrable
  Bootstrap in Planar N=4 SYM Theory}},
  \href{https://arxiv.org/abs/1505.06745}{{\ttfamily 1505.06745}}.

\bibitem{Frolov:2021bwp}
S.~Frolov and A.~Sfondrini, \emph{{Mirror thermodynamic Bethe ansatz for
  AdS3/CFT2}}, \href{https://doi.org/10.1007/JHEP03(2022)138}{\emph{JHEP}
  {\bfseries 03} (2022) 138}
  [\href{https://arxiv.org/abs/2112.08898}{{\ttfamily 2112.08898}}].

\bibitem{Frolov:2021fmj}
S.~Frolov and A.~Sfondrini, \emph{{New dressing factors for AdS3/CFT2}},
  \href{https://doi.org/10.1007/JHEP04(2022)162}{\emph{JHEP} {\bfseries 04}
  (2022) 162} [\href{https://arxiv.org/abs/2112.08896}{{\ttfamily
  2112.08896}}].

\bibitem{OhlssonSax:2018hgc}
O.~Ohlsson~Sax and B.~Stefa\'nski, \emph{{Closed strings and moduli in
  AdS$_{3}$/CFT$_{2}$}},
  \href{https://doi.org/10.1007/JHEP05(2018)101}{\emph{JHEP} {\bfseries 05}
  (2018) 101} [\href{https://arxiv.org/abs/1804.02023}{{\ttfamily
  1804.02023}}].

\bibitem{Kazakov:2004qf}
V.~A. Kazakov, A.~Marshakov, J.~A. Minahan and K.~Zarembo,
  \emph{{Classical/quantum integrability in AdS/CFT}},
  \href{https://doi.org/10.1088/1126-6708/2004/05/024}{\emph{JHEP} {\bfseries
  05} (2004) 024} [\href{https://arxiv.org/abs/hep-th/0402207}{{\ttfamily
  hep-th/0402207}}].

\bibitem{Gromov:2006kve}
N.~Gromov, V.~Kazakov and P.~Vieira, \emph{{Classical limit of Quantum
  Sigma-Models from Bethe Ansatz}},
  \href{https://doi.org/10.22323/1.038.0005}{\emph{PoS} {\bfseries SOLVAY}
  (2006) 005} [\href{https://arxiv.org/abs/hep-th/0703137}{{\ttfamily
  hep-th/0703137}}].

\bibitem{Bombardelli:2018bqz}
D.~Bombardelli, A.~Cavagli\`a, R.~Conti and R.~Tateo, \emph{{Exploring the
  spectrum of planar AdS$_{4}$/CFT$_{3}$ at finite coupling}},
  \href{https://doi.org/10.1007/JHEP04(2018)117}{\emph{JHEP} {\bfseries 04}
  (2018) 117} [\href{https://arxiv.org/abs/1803.04748}{{\ttfamily
  1803.04748}}].

\bibitem{Marboe_2015}
C.~Marboe and D.~Volin, \emph{Quantum spectral curve as a tool for a
  perturbative quantum field theory},
  \href{https://doi.org/10.1016/j.nuclphysb.2015.08.021}{\emph{Nuclear Physics
  B} {\bfseries 899} (2015) 810}.

\bibitem{Marboe_2018}
C.~Marboe and D.~Volin, \emph{The full spectrum of {AdS}5/{CFT}4 i:
  representation theory and one-loop q-system},
  \href{https://doi.org/10.1088/1751-8121/aab34a}{\emph{Journal of Physics A:
  Mathematical and Theoretical} {\bfseries 51} (2018) 165401}.

\bibitem{Marboe:2018ugv}
C.~Marboe and D.~Volin, \emph{{The full spectrum of AdS$_5$/CFT$_4$ II: Weak
  coupling expansion via the quantum spectral curve}},
  \href{https://doi.org/10.1088/1751-8121/abd59c}{\emph{J. Phys. A} {\bfseries
  54} (2021) 055201} [\href{https://arxiv.org/abs/1812.09238}{{\ttfamily
  1812.09238}}].

\bibitem{Apolo:2022fya}
L.~Apolo, A.~Belin, S.~Bintanja, A.~Castro and C.~A. Keller, \emph{{Deforming
  symmetric product orbifolds: a tale of moduli and higher spin currents}},
  \href{https://doi.org/10.1007/JHEP08(2022)159}{\emph{JHEP} {\bfseries 08}
  (2022) 159} [\href{https://arxiv.org/abs/2204.07590}{{\ttfamily
  2204.07590}}].

\bibitem{Kotikov_2004}
A.~Kotikov, L.~Lipatov, A.~Onishchenko and V.~Velizhanin, \emph{Three-loop
  universal anomalous dimension of the wilson operators in n=4 {SUSY}
  yang{\textendash}mills model},
  \href{https://doi.org/10.1016/j.physletb.2004.05.078}{\emph{Physics Letters
  B} {\bfseries 595} (2004) 521}.

\bibitem{Loewy:2002gf}
A.~Loewy and Y.~Oz, \emph{{Large spin strings in $AdS_3$}},
  \href{https://doi.org/10.1016/S0370-2693(03)00196-5}{\emph{Phys. Lett. B}
  {\bfseries 557} (2003) 253}
  [\href{https://arxiv.org/abs/hep-th/0212147}{{\ttfamily hep-th/0212147}}].

\bibitem{David:2014qta}
J.~R. David and A.~Sadhukhan, \emph{{Spinning strings and minimal surfaces in
  $AdS_3$ with mixed 3-form fluxes}},
  \href{https://doi.org/10.1007/JHEP10(2014)049}{\emph{JHEP} {\bfseries 10}
  (2014) 049} [\href{https://arxiv.org/abs/1405.2687}{{\ttfamily 1405.2687}}].

\bibitem{Banerjee:2015qeq}
A.~Banerjee and A.~Sadhukhan, \emph{{Multi-spike strings in AdS$_{3}$ with
  mixed three-form fluxes}},
  \href{https://doi.org/10.1007/JHEP05(2016)083}{\emph{JHEP} {\bfseries 05}
  (2016) 083} [\href{https://arxiv.org/abs/1512.01816}{{\ttfamily
  1512.01816}}].

\bibitem{Alday:2007mf}
L.~F. Alday and J.~M. Maldacena, \emph{{Comments on operators with large
  spin}}, \href{https://doi.org/10.1088/1126-6708/2007/11/019}{\emph{JHEP}
  {\bfseries 11} (2007) 019} [\href{https://arxiv.org/abs/0708.0672}{{\ttfamily
  0708.0672}}].

\bibitem{Gromov:2014bva}
N.~Gromov, F.~Levkovich-Maslyuk, G.~Sizov and S.~Valatka, \emph{{Quantum
  spectral curve at work: from small spin to strong coupling in $ \mathcal{N} $
  = 4 SYM}}, \href{https://doi.org/10.1007/JHEP07(2014)156}{\emph{JHEP}
  {\bfseries 07} (2014) 156} [\href{https://arxiv.org/abs/1402.0871}{{\ttfamily
  1402.0871}}].

\bibitem{Cagnazzo_2012}
A.~Cagnazzo and K.~Zarembo, \emph{B-field in {AdS} 3/{CF} t 2 correspondence
  and integrability},
  \href{https://doi.org/10.1007/jhep11(2012)133}{\emph{Journal of High Energy
  Physics} {\bfseries 2012} (2012) }.

\bibitem{Hoare:2013pma}
B.~Hoare and A.~A. Tseytlin, \emph{{On string theory on $AdS_3 \times S^3
  \times T^4$ with mixed 3-form flux: tree-level S-matrix}},
  \href{https://doi.org/10.1016/j.nuclphysb.2013.05.005}{\emph{Nucl. Phys. B}
  {\bfseries 873} (2013) 682}
  [\href{https://arxiv.org/abs/1303.1037}{{\ttfamily 1303.1037}}].

\bibitem{Hoare:2013ida}
B.~Hoare and A.~A. Tseytlin, \emph{{Massive S-matrix of $AdS_3 \times S^3
  \times T^4$ superstring theory with mixed 3-form flux}},
  \href{https://doi.org/10.1016/j.nuclphysb.2013.04.024}{\emph{Nucl. Phys. B}
  {\bfseries 873} (2013) 395}
  [\href{https://arxiv.org/abs/1304.4099}{{\ttfamily 1304.4099}}].

\bibitem{Lloyd:2014bsa}
T.~Lloyd, O.~Ohlsson~Sax, A.~Sfondrini and B.~Stefa\'nski, Jr., \emph{{The
  complete worldsheet S matrix of superstrings on AdS$_3 \times$ S$^3 \times$
  T$^4$ with mixed three-form flux}},
  \href{https://doi.org/10.1016/j.nuclphysb.2014.12.019}{\emph{Nucl. Phys. B}
  {\bfseries 891} (2015) 570}
  [\href{https://arxiv.org/abs/1410.0866}{{\ttfamily 1410.0866}}].

\bibitem{Leurent_2013}
S.~Leurent and D.~Volin, \emph{Multiple zeta functions and double wrapping in
  planar $n=4$ sym},
  \href{https://doi.org/10.1016/j.nuclphysb.2013.07.020}{\emph{Nuclear Physics
  B} {\bfseries 875} (2013) 757}.

\end{thebibliography}\endgroup

\end{document}